\documentclass[10pt]{article}

\usepackage{amsmath}
\usepackage{amsthm}
\usepackage{amssymb}
\numberwithin{equation}{section}

\theoremstyle{plain}
\newtheorem{proposition}{Proposition}[section]
\newtheorem{corollary}[proposition]{Corollary}
\newtheorem{lemma}[proposition]{Lemma}
\newtheorem{theorem}[proposition]{Theorem}

\theoremstyle{definition}
\newtheorem{definition}[proposition]{Definition}
\newtheorem{example}[proposition]{Example}
\newtheorem{remark}[proposition]{Remark}
\newtheorem{assumption}[proposition]{Assumption}

\newcommand{\ds}{\displaystyle}

\newcommand{\R}{\mathbf{R}}
\newcommand{\wt}[1]{\widetilde{#1}}

\newcommand{\abs}[1]{\left\lvert #1 \right\rvert}

\DeclareMathOperator{\sgn}{sgn}

\DeclareMathOperator*{\res}{res}

\DeclareMathOperator{\linspan}{span}
\DeclareMathOperator{\id}{id}

\begin{document}

\title{Degasperis--Procesi peakons and the discrete cubic string}
\author{%
Hans Lundmark\\
Department of Mathematics\\
Link{\"o}ping University\\
SE-581 83 Link{\"o}ping\\
SWEDEN\\
halun@mai.liu.se
\and
Jacek Szmigielski\\
Department of Mathematics and Statistics\\
University of Saskatchewan\\
106 Wiggins Road\\
Saskatoon, Saskatchewan, S7N 5E6\\
CANADA\\
szmigiel@math.usask.ca
}
\date{March 15, 2005}
\maketitle

\begin{abstract}
  We use an inverse scattering approach to study multi-peakon
  solutions of the Degasperis--Procesi (DP) equation, an integrable PDE
  similar to the Camassa--Holm shallow water equation.
  The spectral problem associated to the DP equation is
  equivalent under a change of variables to what we
  call the cubic string problem,
  which is a third order non-selfadjoint generalization of the
  well-known equation describing the vibrational modes of an
  inhomogeneous string attached at its ends.
  We give two proofs that the eigenvalues of the cubic string are positive
  and simple;
  one using scattering properties of DP peakons,
  and another using the Gantmacher--Krein theory of oscillatory kernels.

  For the discrete cubic string (analogous to a string consisting of
  $n$ point masses) we solve explicitly the inverse spectral problem
  of reconstructing the mass distribution from suitable spectral data,
  and this leads to explicit formulas for the general $n$-peakon
  solution of the DP equation.
  Central to our study of the inverse problem is a peculiar type of
  simultaneous rational approximation of the two Weyl functions of the
  cubic string, similar to classical Pad\'e--Hermite approximation
  but with lower order of approximation and an additional symmetry
  condition instead.
  The results obtained are intriguing and nontrivial generalizations of
  classical facts from the theory of Stieltjes continued fractions
  and orthogonal polynomials.
\end{abstract}







To appear in \emph{International Mathematics Research Papers}.

\section{Introduction}

The PDE
\begin{equation}
  \label{eq:family}
  u_t - u_{xxt} + (b+1) u u_x = b u_x u_{xx} + u u_{xxx},
  \quad (x,t) \in \R^2,
\end{equation}
is known to be completely integrable if and only if $b=2$ or $b=3$.
The case $b=2$ is the well-known
Camassa--Holm (CH) shallow water equation \cite{ch}.
The case $b=3$ is the Degasperis--Procesi (DP) equation,
found by Degasperis and Procesi \cite{dp} to pass the
necessary (but not sufficient) test of asymptotic integrability,
and recently shown by Degasperis, Holm and Hone \cite{dhh1,dhh2}
to be integrable indeed.

All equations in the family \eqref{eq:family}
admit (in a weak sense) a type of non-smooth solutions called
\emph{multi-peakons}
(peakon = peaked soliton).
These take the form of a train of peak-shaped interacting waves,
\begin{equation}
  u(x,t) = \sum_{i=1}^n m_i(t) \, e^{-\abs{x-x_i(t)}},
\end{equation}
where the positions $x_i(t)$ and the heights $m_i(t)$
are determined by the following system of nonlinear ODEs:
\begin{equation}
  \label{eq:general-peakonODE}
    \begin{split}
    \dot{x}_k &= \sum_{i=1}^n m_i e^{-\abs{x_k-x_i}},
    \\
    \dot{m}_k &= (b-1) \sum_{i=1}^n m_k m_i \sgn(x_k-x_i) \, e^{-\abs{x_k-x_i}},
  \end{split}
\end{equation}
for $k=1,\ldots,n$.
Here we use the convention that $\sgn 0 = 0$,
and dots denote $\frac{d}{dt}$ as usual.
Throughout the paper, $n$ will be fixed but arbitrary.

In the integrable CH and DP cases,
explicit solution formulas for $x_i(t)$ and $m_i(t)$
can be found using inverse scattering techniques,
which makes it possible to analyze the peakon interactions in great detail.
This was shown for the CH case $b=2$
by Beals, Sattinger and Szmigielski \cite{bss-stieltjes,bss-moment}.
The DP case $b=3$ was briefly treated in a short note by the present authors
\cite{ls-invprob},
where we outlined a recursive procedure to determine
solution formulas for any given $n$.
It is our purpose here to continue this study in more detail.
Among other things,
we will give the promised closed form $n$-peakon solution of
the DP equation
(see Corollary~\ref{cor:n-peakon-solution}).

This paper is divided into two major parts.
In Section~\ref{sec:peakons} we study the peakon solutions of
the DP equation.
First we prove certain scattering properties of the peakons,
using only elementary methods and the governing equations themselves.
Then we describe the inverse scattering procedure,
based on the Lax pair given in \cite{dhh1}.
The spectral problem associated to the DP equation is of third order
and non-selfadjoint.
To prove that the spectrum is nevertheless real and simple,
as well as some other related facts that we need,
we make extensive use of the scattering properties derived earlier.
We state the resulting formulas for the general $n$-peakon solution,
and extract precise asymptotic properties as \mbox{$t\to\pm\infty$}.

Sections~\ref{sec:cubic-string} and~\ref{sec:inverse-cubic}
are devoted to the forward and inverse spectral problems for the
\emph{discrete cubic string},
our main tool for proving the explicit $n$-peakon formulas,
but also a very interesting problem in its own right from the
point of view of operator theory.
By the forward cubic string problem,
we mean the following third order spectral problem:
for a given function $g(y)\ge 0$,
determine the eigenvalues $z$ such that nontrivial eigenfunctions
$\phi(y)$ exist satisfying
\begin{gather*}
  -\phi_{yyy}(y) = z \, g(y) \, \phi(y)
  \quad\text{for $y \in (-1,1)$},
  \\
  \phi(-1) = \phi_y(-1) = 0,
  \qquad
  \phi(1) = 0.
\end{gather*}
This spectral problem, which is equivalent under a change of variables
to the one appearing in the DP Lax pair,
is a non-selfadjoint generalization of the well-known (selfadjoint)
string equation
\begin{gather*}
  \phi_{yy}(y) = z \, g(y) \, \phi(y)
  \quad\text{for $y \in (-1,1)$},
  \\
  \phi(-1) = 0,
  \qquad
  \phi(1) = 0,
\end{gather*}
describing the vibrational modes of a string with
nonhomogeneous mass density $g(y)$, attached at the ends $y=\pm 1$.
The discrete case arises when $g=\sum_{i=1}^n g_i \, \delta_{y_i}$
is a discrete measure (``point masses $g_i$ at the positions $y_i$''),
so that the eigenfunctions are piecewise linear in~$y$ for
the ordinary string and piecewise quadratic for the cubic string.

From what we proved about peakon scattering, it follows that
the eigenvalues of the discrete cubic string are positive and simple.
We also provide an independent proof of this,
valid not only in the discrete case,
using the theory of oscillatory kernels developed by Gantmacher and Krein \cite{gantmacher-krein}.

The discrete (ordinary) string plays a crucial role in finding the general
$n$-peakon solution for the CH equation \cite{bss-stieltjes,bss-moment}.
The inverse spectral problem consists of determining the positions $y_i$ and
masses $g_i$ given the eigenfrequencies and suitable additional information
about the eigenfunctions
(encoded in the \emph{spectral measure} of the string,
or equivalently in its Stieltjes transform, henceforth called
the \emph{Weyl function}).
The solution, found by Krein (and outlined in Appendix~\ref{sec:app-string}
for the reader's convenience),
involves Stieltjes continued fractions,
a subject with extremely rich connections to various mathematical areas such
as the classical moment problem, Pad\'e approximation,
and orthogonal polynomials.

The corresponding inverse problem for the discrete cubic string,
which we solve in Section~\ref{sec:inverse-cubic},
leads to new and very intriguing generalizations of much of this.
There are two Weyl function, one of which is unexpectedly determined by
the other (the proof of this ``static'' fact uses the isospectral
deformation of the cubic string induced by the DP dynamics).
These two Weyl functions admit simultaneous approximation by
rational functions with a common denominator,
similar to classical Pad\'{e}--Hermite approximation of two
(independently chosen) functions,
but with lower order of approximation and instead an additional symmetry
condition,
and also similar to the Pad\'e approximation provided by the convergents
of the Stieltjes continued fraction in the ordinary string case.
The study of this new type of approximation problem is the key to the
solution of the inverse spectral problem.
The coefficients in the common denominator of the rational approximants
are given by ratios of determinants
involving certain moment-like \emph{double} integrals of a spectral measure.
The evaluation of these determinants is significantly more
difficult than for the classical case,
where the well-known Hankel determinants of moments
of the spectral measure appear instead.

It may be expected that Krein's solution of the inverse spectral problem
in the case of a string with a general mass distribution
also admits a generalization to the cubic string,
but we have not endeavored to address that question here.

\section{Peakon solutions of the DP equation}
\label{sec:peakons}

\subsection{The governing equations}

We begin by summarizing some basic facts about the
Degasperis--Procesi (DP) equation
\begin{equation}
  \label{eq:DPsingle}
  u_t - u_{xxt} + 4 u u_x = 3 u_x u_{xx} + u u_{xxx}.
\end{equation}
It is advantageous to write it as a system,
\begin{subequations}
  \label{eq:DP}
  \begin{align}
    0 &= m_t + m_x u + 3 m u_x, \label{eq:DP1}
    \\
    m &= u - u_{xx}.
    \label{eq:DP2}
  \end{align}
\end{subequations}
The function $m(x,t)$ can be thought of as a momentum-like quantity,
although we are not aware of any actual physical interpretation of the
DP equation.
The $n$-peakon solution
\begin{equation}
  \label{eq:peakons}
  u(x,t) = \sum_{i=1}^n m_i(t) \, e^{-\abs{x-x_i(t)}}
\end{equation}
 arises when $m$ is not a function but a finite discrete measure,
\begin{equation}
  \label{eq:measure-m}
  m(x,t) = 2 \sum_{i=1}^n m_i(t) \, \delta\bigl(x-x_i(t)\bigr).
\end{equation}
Then \eqref{eq:DP2} is satisfied by construction,
while \eqref{eq:DP1} is satisfied in a weak sense if and only if
the positions $x_1(t),\ldots,x_n(t)$ and the heights $m_1(t),\ldots,m_n(t)$
satisfy the following system, which was given in \cite{dhh1}
and is one of our fundamental objects of study here:
\begin{equation}
  \label{eq:peakon-dynamics}
  \begin{split}
    \dot{x}_k &= \sum_{i=1}^n m_i e^{-\abs{x_k-x_i}},
    \\
    \dot{m}_k &= 2 \sum_{i=1}^n m_k m_i \sgn(x_k-x_i) \, e^{-\abs{x_k-x_i}},
  \end{split}
\end{equation}
for $k=1,\ldots,n$.

\subsection{Examples of explicit solutions}

Finding the solution of \eqref{eq:peakon-dynamics}
in the case $n=1$ is trivial:
$m_1 = \text{constant}$ and $x_1=x_1(0)+m_1 t$.
We can write this in a way that anticipates the general pattern:
\begin{equation}
  \label{eq:onepeakon}
  m_1(t) = \frac{b_1^2}{\lambda_1 b_1^2},
  \qquad
  x_1(t) = \log b_1,
\end{equation}
where $b_1(t)=b_1(0) e^{t/\lambda_1}$,
and the constants
$\lambda_1 = 1/m_1(0)$
and
$b_1(0) = e^{x_1(0)}$
are determined by initial conditions.

Also when $n=2$, the equations can be integrated by
straightforward methods if one changes to variables
$x_1 \pm x_2$ and $m_1 \pm m_2$ as discussed in \cite{dhh1}.
In our notation, the solution takes the form
\begin{equation}
  \label{eq:twopeakon}
  \begin{split}
    x_1(t) &= \log \frac{\frac{(\lambda_1-\lambda_2)^2}{\lambda_1+\lambda_2}b_1 b_2}{\lambda_1 b_1 + \lambda_2 b_2},
    \\
    x_2(t) &= \log (b_1+b_2),
    \\
    m_1(t) &= \frac{(\lambda_1 b_1 + \lambda_2 b_2)^2}{\lambda_1 \lambda_2 \left( \lambda_1 b_1^2 + \lambda_2 b_2^2 + \frac{4 \lambda_1 \lambda_2}{\lambda_1+\lambda_2}b_1 b_2 \right)},
    \\
    m_2(t) &= \frac{(b_1+b_2)^2}{\lambda_1 b_1^2 + \lambda_2 b_2^2 + \frac{4 \lambda_1 \lambda_2}{\lambda_1+\lambda_2}b_1 b_2},
  \end{split}
\end{equation}
where $b_k(t)=b_k(0) e^{t/\lambda_k}$,
and the constants
$\lambda_1$, $\lambda_2$, $b_1(0)$, $b_2(0)$
are determined by initial conditions.

For $n\ge 3$ the peakon equations become considerably more involved,
and so also the explicit solution formulas,
if written out in full.
However, if one introduces suitable notation for certain symmetric functions
of $b_k$'s and $\lambda_k$'s,
the formulas become very succinct;
see
Definitions~\ref{def:notation-galore} and~\ref{def:UVW} for notation,
Corollary~\ref{cor:n-peakon-solution} for the general statement
and Example~\ref{ex:threepeakon} for the three-peakon solution
written out in detail.

\begin{remark}
  To write the formulas for $x_1(t)$ and $x_2(t)$ in
  \eqref{eq:twopeakon} in the form given in \cite{dhh1}, substitute
  \begin{gather*}
    \lambda_1 = \frac{1}{c_1},
    \qquad
    \lambda_2 = \frac{1}{c_2},
    \qquad
    P=c_1+c_2,
    \\
    b_1(0) = \frac{\sqrt{\Gamma}}{2(c_1-c_2)},
    \qquad
    b_2(0) = \frac{2c_2 P}{(c_1-c_2) \sqrt{\Gamma}}.
  \end{gather*}
  There is a small error in \cite{dhh1}: the term $4 c_2 P$ in
  their equation (6.8) should be $4 c_1 P$, and vice versa in (6.9).
  However, their phase shift formulas (6.10) and (6.11) are correct,
  and consistent with the corrected (6.8) and (6.9).
\end{remark}

\subsection{A priori estimates}

In this section we derive some useful information about the peakon dynamics
directly from the governing equations \eqref{eq:peakon-dynamics}.
Although this is instructive in itself,
our primary reason for doing this is that we will use this information
in order to show that certain eigenvalues $\lambda_k$ that appear later
in the inverse scattering analysis are real and simple.

There is a distinction between peakons ($m_k>0$) and antipeakons ($m_k<0$).
As indicated by the formulas for $n=1$ and $n=2$ above,
peakons move to the right and antipeakons to the left.
If both are present, the system \eqref{eq:peakon-dynamics}
blows up after a finite time when the first
collision between a peakon and an antipeakon occurs (their corresponding
$m_k$'s diverge to $+\infty$ and $-\infty$, respectively).
In the case of the CH equation, $u(x,t)$ can nevertheless be
continued in a natural way beyond the time of collision \cite{bss-moment}.
The behaviour of the DP equation seems to be more complicated in this respect.
In this paper we will concentrate on the pure peakon case (all $m_k>0$),
except for Section~\ref{sec:antipeakons} where
we will make a few comments about antipeakons.

\begin{assumption}
  \label{ass:ordering}
  Except where otherwise stated, we assume the following:
  \begin{enumerate}
  \item Pure peakons: $m_k(t)>0$ for $k=1,\ldots,n$.
  \item Ordering: $x_1(t)<x_2(t)<\cdots<x_n(t)$.
  \end{enumerate}
\end{assumption}

We call the set where Assumption~\ref{ass:ordering} holds the
 ``pure peakon sector'' of $\R^{2n}$ and we denote it by $\mathcal{P}$.
In other words,
\begin{equation}
  \label{eq:peakon-sector}
  \mathcal{P} =
  \bigl\{ (x,m)\in\R^{2n} \;:\; x_1<\cdots<x_n, \; \text{all $m_i>0$} \bigr\}.
\end{equation}
For later convenience, we also introduce the definitions
\begin{equation}
  \label{eq:x0}
  x_0=-\infty \quad\text{and}\quad x_{n+1}=\infty,
\end{equation}
with natural interpretations like $e^{x_0}=0$ and $\tanh x_0=-1$.

\begin{proposition}[Global pure peakon solution]
  \label{prop:global-solution}
  Given initial conditions consistent with
  Assumption~\ref{ass:ordering} at some time $t=t_0$,
  there exists a unique solution to \eqref{eq:peakon-dynamics}
  defined for all $t\in \R$
  and satisfying these initial conditions.
  There are constants $C$ and $D$ such that
  $m_k(t) > C > 0$ and $x_{k+1}(t)-x_k(t) > D > 0$
  for all $k$ and all $t\in\R$.
  In particular, the solution remains in the interior of $\mathcal{P}$ for
  all~$t\in\R$.
\end{proposition}

\begin{proof}
  The right-hand side of \eqref{eq:peakon-dynamics} is clearly
  uniformly Lipschitz continuous as a function of $x_k$'s and $m_k$'s
  on any compact subset of $\mathcal{P}$, and thus the local existence
  and uniqueness follows from the Picard--Lindel\"{o}f theorem.
  Moreover, since the right-hand side of \eqref{eq:peakon-dynamics} is
  smooth on $\mathcal{P}$, the solution is smooth on the domain of its
  existence. To show global existance, we need to rule out for the
  solution to: (1) intersect in finite time the boundary of
  $\mathcal{P}$, (2) blow up in finite time.

  It can be verified directly from the equations \eqref{eq:peakon-dynamics}
  that
  $M_1=m_1+\ldots+m_n$
  and
  $M_n=
  \Bigl(
  \prod_{k=1}^n m_k
  \Bigr)
  \Bigl(
  \prod_{k=1}^{n-1} (1-e^{x_k-x_{k+1}})^2
  \Bigr)$
  are constants of motion
  as long as the ordering assumption is satisfied,
  which is at least true for all~$t$ in some open interval $T$ containing~$t_0$.
  (In Theorem~\ref{thm:constants-of-motion} we show that $M_1$ and $M_n$
  belong to a set of $n$ constants of motion $M_1,\ldots,M_n$.)

  The pure peakons assumption implies that $M_1$ and $M_n$ are positive,
  and that $m_k(t) < M_1$ for all~$k$ and all $t\in T$.
  Using this estimate on all but one factor $m_k$ in $M_n$,
  and noting that each factor in the second product is $\le 1$
  due to the ordering assumption,
  we get $M_n < m_k(t) M_1^{n-1} 1^{n-1}$.
  Hence $m_k(t) > M_n / M_1^{n-1} > 0$ for all~$k$ and all $t\in T$,
  so the pure peakons assumption holds in~$T$.
  Since by definition of $T$ the ordering assumption holds there, we
  conclude that both the pure peakon and the ordering assumption hold
  in $T$.
  Similarly,
  $(1-e^{x_k-x_{k+1}})^2 > M_n / ( M_1^n 1^{n-2}) =: c \in (0,1)$,
  hence
  $x_{k+1}(t) - x_k(t) > - \log(1-c) > 0$ for each~$k$.
  Since the distances between peakons are bounded from below by a
  strictly positive constant,
  the peakons cannot change places. Thus for no finite time can
   peakons reach the boundary of $\mathcal{P}$.

  Finally, no blow-up of
  of solutions in finite time is admitted because by the second equation
  \eqref{eq:peakon-dynamics}, and with the help of
  $0<M_n / M_1^{n-1} <m_k< M_1$, we get
  $M_n / M_1^{n-1}<\dot{x}_j < M_1$.
  Thus $(M_n / M_1^{n-1}) t < x_j - x_j(0) < M_1 t$, implying that
  the solution stays bounded at any finite time and
  the ordering assumption holds for all~$t$ (that is, $T=\R$).
\end{proof}

By direct analysis of \eqref{eq:peakon-dynamics}
we obtain the following theorem
which shows that the peakons scatter and asymptotically
behave like free particles travelling with distinct velocities.
This behaviour in mechanical systems, integrable by the Lax method,
was first observed and analyzed by J. Moser in \cite {M}.

The results of the next theorem will be superseded by more precise ones
in Section~\ref{sec:asymptotics} when we have at our disposal
the explicit formulas for the $n$-peakon solution.
However, we will use these \emph{a priori} estimates in order to show that
certain eigenvalues $\lambda_1,\ldots,\lambda_n$
appearing in the solution formulas must be real and distinct.

\begin{theorem}[Asymptotics]
  \label{thm:apriori-asymptotics}
  Let $\left\{ x_k(t), m_k(t) \right\}$ be a solution to
  \eqref{eq:peakon-dynamics}
  satisfying Assumption~\ref{ass:ordering}.
  Then the following claims are true:
  \begin{enumerate}
  \item The limits $m_k(\pm\infty) := \ds\lim_{t\to\pm\infty} m_k(t)$
    exist and are finite.
  \item Peakons travel to the right and scatter;
    i.e., $\dot{x}_k(t) >0$,
    and
    $x_{k+1}(t) - x_k(t) \to \infty$ as \mbox{$t\to\pm\infty$}.
  \item $\ds\lim_{t\to\pm\infty} \dot{x}_k(t) = m_k(\pm\infty)$,
    and
    $x_k(t) = m_k(\pm\infty) \, t + O(1)$ as \mbox{$t\to\pm\infty$}.
  \item $0 < m_1(\infty) < m_2(\infty) < \cdots < m_n(\infty)$
    and
    $0 < m_n(-\infty) < \cdots < m_2(-\infty) < m_1(-\infty)$.
  \end{enumerate}
\end{theorem}

\begin{proof}
  We will only prove the statements concerning \mbox{$t\to+\infty$},
  as the case \mbox{$t\to-\infty$} is completely analogous.
  To improve readability,
  let $d_{ij}=\abs{x_i-x_j}$ denote the distance between the
  $i$th peakon and the $j$th.
  If we impose the restriction that the smaller index always be written first
  ($i<j$),
  we can remove the absolute value signs by
  Assumption~\ref{ass:ordering} and Proposition~\ref{prop:global-solution}:
  $d_{ij} = x_j - x_i > 0$.
  Then the peakon equations \eqref{eq:peakon-dynamics} read
  \begin{equation}
    \label{eq:peakon-dynamics again}
    \begin{split}
      \dot{x}_k &=
      \left( \sum_{i<k} m_i \, e^{-d_{ik}} \right) + m_k + \left( \sum_{i>k} m_i \, e^{-d_{ki}} \right),
      \\
      \frac{\dot{m}_k}{2 m_k} &=
      \left( \sum_{i<k} m_i \, e^{-d_{ik}} \right) - \left( \sum_{i>k} m_i \, e^{-d_{ki}} \right).
    \end{split}
  \end{equation}

  We prove claims~1 and~2 together.
  To begin with, it follows from \eqref{eq:peakon-dynamics again} that
  $\dot{x}_k > 0$ since all $m_k > 0$.
  Now we proceed by induction on $k$,
  starting with $n$ and descending to $1$.
  The equation for $m_n$ is
  \begin{equation*}
    \frac{\dot{m}_n}{2m_n} = \sum_{i=1}^{n-1} m_i \, e^{-d_{in}} > 0,
  \end{equation*}
  so $m_n$ is increasing, and since $m_n$ is bounded from above by
  the constant of motion $M_1=\sum_1^n m_i$,
  it follows that $\lim_{t\to\infty} m_n$ exists and is finite.
  Hence the integral in
  \begin{equation*}
    m_n(t) =
    m_n(0) \exp \left(
      \int_0^t 2\sum_{i=1}^{n-1} m_i(s) \, e^{-d_{in}(s)}\, ds
    \right),
  \end{equation*}
  is convergent.
  Since the $m_i$'s are positive and bounded away from zero,
  this implies that
  \begin{equation*}
    \int_0^{\infty} e^{-d_{in}(s)}\, ds < \infty, \qquad  1 \le i \le n-1.
  \end{equation*}
  The derivative of the integrand $e^{-d_{in}(t)}$ is bounded:
  $\abs{\dot{x}_k} \le \sum_1^n m_i = M_1$
  by \eqref{eq:peakon-dynamics again}, so
  \begin{equation*}
    \abs{\frac{d}{dt} e^{-d_{in}(t)}}
    = \abs{(\dot{x}_n-\dot{x}_i) e^{-d_{in}}}
    \le \abs{\dot{x}_n} + \abs{\dot{x}_i}
    \le 2 M_1.
  \end{equation*}
  This means that $e^{-d_{in}(t)}$ must vanish as \mbox{$t\to\infty$},
  otherwise the integral would be divergent.
  (Suppose $u(t)>0$,
  $\abs{\dot{u}(t)} \le C$,
  and $u$ does \emph{not} vanish as \mbox{$t\to\infty$};
  say $\limsup_{t\to\infty} u(t) = \epsilon > 0$.
  Then there is an increasing sequence of points $t_j \to\infty$
  (which can be taken to lie more than $\epsilon/4C$ apart)
  such that $u(t_j) > \epsilon/2$,
  and hence $u(t)> \epsilon/4$ in the interval
  $[t_j,t_j+\epsilon/4C]$ because of the bound on $\dot{u}$.
  This forces
  $\int_0^{\infty} u(s)\,ds = \infty$.)
  Consequently,
  the $n$th peakon scatters away from the ones to its left:
  \begin{equation}
    \lim_{t\to\infty} d_{in}(t) = \infty,
    \qquad
    i=1,\ldots,n-1.
  \end{equation}

  We have now proved the base step of the induction.
  When considering the $k$th peakon,
  we make the following induction hypothesis
  about the peakons to its right:
  for all $j>k$,
  \begin{enumerate}
  \item[i.] $m_j(\infty) = \ds\lim_{t\to\infty} m_j(t)$ exists and is finite.
  \item[ii.] The $j$th peakon scatters away from the ones to its left:
    $\int_0^{\infty} e^{-d_{ij}(s)} < \infty$
    and $\ds\lim_{t\to \infty} d_{ij}(t)=\infty$
    for $i=1,\ldots,j-1$.
  \end{enumerate}
  We need to show that the same holds also for $j=k$.
  Integration of the equation for $m_k$ in
  \eqref{eq:peakon-dynamics again} yields
  \begin{multline*}
    m_k(t)
    \exp \left(
      2\int_0^{t}\sum _{j>k} m_j(s) \, e^{-d_{kj}(s)}\, ds
    \right)
    =\\
    =m_k(0)
    \exp \left(
      2\int_0^{t}\sum _{i<k} m_i(s) \, e^{-d_{ik}(s)}\, ds
    \right).
  \end{multline*}
  Since the right-hand side is an increasing function
  of~$t$,
  we know that its limit exists,
  even though it might be infinite.
  By the induction hypothesis the integral
  on the left-hand side has a finite limit.
  We conclude that
  $m_k(\infty) = \lim_{t\to \infty} m_k(t)$ exists,
  and this limit is finite because $m_k$ is bounded.
  Hence the limit of the right-hand side is in fact finite,
  which, using the same arguments as in the
  base step of the induction, implies
  $\int_0^{\infty} e^{-d_{ik}(s)} \, ds < \infty$
  and
  $\lim_{t\to \infty} d_{ik}(t)=\infty$
  for $i=1,\ldots,k-1$;
  that is, the $k$th peakon scatters away from the ones to its left.
  This concludes the proof of claims~1 and~2.

  The scattering property just proved implies that all terms
  in the equation for $\dot{x}_k$ in \eqref{eq:peakon-dynamics again}
  vanish as \mbox{$t\to\infty$}, except the lone $m_k$.
  Consequently, by l'Hospital's rule,
  $\lim_{t\to \infty} x_k(t)/t
  =\lim_{t\to \infty} \dot {x}_k/1
  =m_k(\infty)$,
  which doesn't quite prove claim~3 yet,
  since at this stage we can only draw the following weaker conclusion:
  \begin{equation}
    \label{eq:asymptotically-free}
    x_k(t) = m_k(\infty) \, t + o(t)
    \qquad\text{as $t\to\infty$}.
  \end{equation}
  We will improve the remainder to $O(1)$ after using
  \eqref{eq:asymptotically-free} to prove claim~4.

  Proposition~\ref{prop:global-solution} tells us
  that $m_1$ is bounded away from zero,
  hence $0 < m_1(\infty)$,
  and that the ordering $x_1<\cdots<x_n$ is preserved,
  hence (in order not to contradict \eqref{eq:asymptotically-free})
  \begin{equation*}
    m_1(\infty) \le m_2(\infty) \le \cdots \le m_n(\infty).
  \end{equation*}
  To prove claim~4,
  it remains to show that these inequalities are actually strict.
  Suppose, to derive a contradiction, that not all $m_k(\infty)$ are distinct.
  Then there is a sequence of two or more
  consecutive $m_k(\infty)$ that are equal, say
  \begin{equation}
    \label{eq:contradiction m}
    m_{a-1}(\infty) < m_a(\infty) = \cdots = m_b(\infty) < m_{b+1}(\infty)
  \end{equation}
  for some $a<b$.
  (If $a=1$, then $m_{a-1}(\infty)$ is not present here,
  and similarly $m_{b+1}(\infty)$ is not present if $b=n$.)
  We will show that this implies the existence of a $t_0$ such that
  $\dot{m}_a(t) < 0$  and $\dot{m}_b(t) > 0$ for all $t\ge t_0$.
  Then we will show that one can find $t_1 > t_0$ such that
  $m_a(t_1) < m_b(t_1)$.
  Since $m_a$ will decrease and $m_b$ will increase as $t$ grows,
  it follows that
  $m_a(\infty) < m_b(\infty)$,
  which is the desired contradiction.

  The case $a=1$ is trivial, since it is clear from
  \eqref{eq:peakon-dynamics again} that $m_1(t)$ is decreasing for all~$t$.
  So assume $2 \le a < n$ and rewrite the equation for $m_a$
  by factoring out the exponential in the last positive term ($i=a-1$):
  \begin{equation*}
    \nonumber
    \frac{\dot{m}_a}{2 m_a} =
    e^{-d_{a-1,a}} \left[
      \left(
        \sum_{i=1}^{a-2} m_i \, e^{-d_{i,a-1}}
      \right)
      + m_{a-1}
      - \left(
        \sum_{i={a+1}}^n m_i \, e^{-d_{a,i}+d_{a-1,a}}
      \right)
    \right].
  \end{equation*}
  Under the assumption \eqref{eq:contradiction m},
  equation \eqref{eq:asymptotically-free} implies that
  $d_{a-1,a} = ct+o(t)$ with $c = m_a(\infty)-m_{a-1}(\infty) > 0$,
  and $d_{a,a+1}=o(t)$ (since $m_a(\infty)=m_{a+1}(\infty)$).
  Hence the exponent in the term corresponding to $i=a+1$
  is $ct+o(t)$,
  which,
  since the factor $m_{a+1}$ in front of the exponential
  is bounded away from zero,
  means that the second sum tends to infinity as \mbox{$t\to\infty$}.
  The first sum tends to zero by scattering
  (unless $a=2$ in which case it is empty),
  and $m_{a-1}$ is bounded.
  Consequently,
  the right-hand side is negative for all sufficiently large~$t$,
  so there is a $t_a$ such that $m_a(t)$ is decreasing for $t \ge t_a$.

  Similarly, $m_n(t)$ is increasing so $b=n$ is a trivial case.
  If $1 < b \le n-1$,
  rewrite the equation for $m_b$
  by factoring out the exponential in the
  first negative term ($i=b+1$):
  \begin{equation*}
    \nonumber
    \frac{\dot{m}_b}{2 m_b} =
    e^{-d_{b,b+1}} \left[
      \left(
        \sum_{i=1}^{b-1} m_i \, e^{-d_{ib}+d_{b,b+1}}
      \right)
      - m_{b+1}
      - \left(
        \sum_{i={b+2}}^n m_i \, e^{-d_{b+1,i}}
      \right)
    \right].
  \end{equation*}
  The right-hand side is positive for large~$t$,
  since the second sum is either empty (if $b=n-1$)
  or tends to zero by scattering, $m_{b+1}$ is bounded,
  and the term corresponding to $i=b-1$ in the first sum tends to infinity
  because its exponent is $ct+o(t)$ with $c=m_{b+1}(\infty)-m_b(\infty)>0$.
  Hence $m_b(t)$ is increasing for $t\ge t_b$, say.
  Now simply let $t_0 = \max(t_a,t_b)$. Then
  $\dot{m}_a(t) < 0$ and $\dot{m}_b(t) > 0$ for all $t\ge t_0$,
  as was to be shown.

  Finally, observe that
  \begin{equation*}
    \dot{d}_{ab} = \dot{x}_b - \dot{x}_a =
    m_b - m_a
    + \left( \sum_{i\ne b} m_i e^{-\abs{x_i-x_b}} \right)
    - \left( \sum_{i\ne a} m_i e^{-\abs{x_i-x_a}} \right).
  \end{equation*}
  The integral of this from $t=0$ to $\infty$ diverges,
  since $d_{ab}\to\infty$ by claim~2.
  However, the contributions from the sums in parentheses are finite
  by the proof of claim~2.
  This means that
  \begin{equation*}
    \int_0^{\infty} \bigl( m_b(t) - m_a(t) \bigr) \, dt = +\infty,
  \end{equation*}
  which cannot hold if the integrand is nonpositive for all $t \ge t_0$.
  This shows that there is a $t_1>t_0$ such that $m_b(t_1) - m_a(t_1) > 0$,
  which completes the proof of claim~4.

  Now it only remains to finish the proof of claim~3 by
  sharpening the estimate $o(t)$ in \eqref{eq:asymptotically-free}
  to $O(1)$.
  It will be convenient here to put the absolute value signs back
  in $d_{ij}=\abs{\dot{x}_i-\dot{x}_j}$ and remove the restriction
  $i<j$.
  Let $c_1 = \frac12 \min_{i \ne j} \abs{m_i(\infty)-m_j(\infty)}$.
  By claim~4 we know that $c_1>0$ and then
  \eqref{eq:asymptotically-free} shows that there is a $c_2>0$
  such that on the interval $t>0$, say, the uniform estimate
  \begin{equation}
    \label{eq:exp decay}
    m_i(t) \, e^{-d_{ij}(t)} < c_2 \, e^{-c_1 t}
    \qquad
    (i\ne j)
  \end{equation}
  holds.
  From the equations of motion \eqref{eq:peakon-dynamics again}
  we get
  \begin{multline*}
    x_k(t) - m_k(\infty) t
    = x_k(0) + \int_0^t \bigl( \dot{x}_k(s) - m_k(\infty) \bigr) \, ds =
    \\
    = x_k(0) + \left(
      \sum_{i\ne k} \int_0^t m_i(s) \, e^{-d_{ik}(s)} \, ds
    \right)
    + \int_0^t \bigl( m_k(s) - m_k(\infty) \bigr) \, ds.
  \end{multline*}
  The sum in brackets is bounded as \mbox{$t\to\infty$},
  since the integrals are convergent.
  (We saw this already earlier in this proof, but with the estimate
  \eqref{eq:exp decay} we can even see that they converge exponentially fast.)
  So we need only prove that the last integral is convergent.
  The expression for $\dot{m}_k/2m_k$ in \eqref{eq:peakon-dynamics again}
  contains $n-1$ terms, all of the form \eqref{eq:exp decay},
  so we obtain
  \begin{multline*}
    \abs{m_k(s) - m_k(\infty)}
    \le \int_s^{\infty} \abs{\dot{m}_k(\xi)}\, d\xi
    \le
    \\
    \le \int_s^{\infty} 2 M_1 (n-1) c_2 \, e^{-c_1 \xi} \, d\xi
    = 2 M_1 (n-1) \frac{c_2}{c_1} \, e^{-c_1 s},
  \end{multline*}
  and the integral of the right-hand side from $s=0$ to $\infty$
  is convergent.
  Thus $\int_0^{\infty} \bigl( m_k(s) - m_k(\infty) \bigr) \, ds$
  is absolutely convergent, and the proof is finished.
\end{proof}

\subsection{Lax pair}
\label{sec:laxpair}

The complete integrability of the DP equation manifests itself in the
existence of an associated spectral problem, as is usually the case in
soliton theory.
It was shown in \cite{dhh1}
that the DP equation is the compatibility condition for the following
linear system for the \emph{wavefunction} $\psi(x,t;z)$:
\begin{subequations}
  \label{eq:lax}
  \begin{align}
    \label{eq:lax1}
    (\partial_x - \partial_x^3) \psi &= z \, m\psi, \\
    \label{eq:lax2}
    \psi_t &= \left[ z^{-1} (1-\partial_x^2) + u_x - u \partial_x \right] \psi.
  \end{align}
\end{subequations}
(The term $z^{-1} (1-\partial_x^2)$ could be replaced by
$z^{-1} (c-\partial_x^2)$
where $c$ is an arbitrary constant,
but for our purposes $c=1$ is the appropriate choice.)

Indeed, with
$L=\partial_x^3 - \partial_x + zm$
and
$A=z^{-1} (1-\partial_x^2) + u_x - u \partial_x$,
the generalized Lax equation
\begin{equation}
  \label{eq:generalized-lax}
  L_t = [A,L] - 3 u_x L
\end{equation}
is equivalent to
\begin{equation}
  \label{eq:almostDP}
  m_t + m_x u + 3 m u_x = 0,
  \qquad
  m_x = (u-u_{xx})_x.
\end{equation}
Clearly \eqref{eq:almostDP} is satisfied by solutions of the DP equation
\eqref{eq:DP}.
Conversely, any solution of \eqref{eq:almostDP} such that
$m$ and $u-u_{xx}$ vanish as $\abs{x}\to\infty$
(as is certainly the case for peakons)
also satisfies the DP equation.

Out of all possible solutions to \eqref{eq:lax},
we distinguish the following one in particular:

\begin{definition}[DP wavefunction]
  \label{def:psi-general}
  To a given solution $u(x,t)$, $m(x,t)$ of the DP equation \eqref{eq:DP},
  and to a given complex number $z$,
  we associate the corresponding wavefunction $\psi(x,t;z)$ satisfying
  \eqref{eq:lax} and the additional asymptotic condition
  \begin{equation}
    \label{eq:wavefcn-left}
    \psi(x,t;z) \sim e^x,
    \qquad
    \text{as $x\to-\infty$}.
  \end{equation}
  (It is not hard to see that \eqref{eq:wavefcn-left}
  is consistent with \eqref{eq:lax}
  if $u$ and its derivatives decay sufficiently fast as
  $\abs{x}\to\infty$;
  we show the details for the peakon case
  in Section~\ref{sec:timedep-psi} below.)
\end{definition}

\subsection{The peakon wavefunction $\psi(x,t;z)$}
\label{sec:psi}

For peakon solutions,
when $m=2\sum_1^n m_i \delta_{x_i}$
is a discrete measure,
the wavefunction $\psi$ of Definition~\ref{def:psi-general}
can be written down quite explicitly.
Since in this case the right-hand side of \eqref{eq:lax1}
equals zero away from the points $x_i$,
the wavefunction is a solution of $(\partial_x^3-\partial_x)\psi=0$
in each interval $(x_k,x_{k+1})$.
In other words, $\psi$ is piecewise given by the expression
\begin{equation}
  \label{eq:piecewise-psi}
  \begin{split}
    \psi(x,t;z) &= A_k(t;z) \, e^x + B_k(t;z) + C_k(t;z) \, e^{-x},
    \\
    x &\in \bigl(x_k(t),x_{k+1}(t) \bigr)
    \qquad (k=0,\dots,n).
  \end{split}
\end{equation}
(Recall our convention that $x_0=-\infty$ and $x_{n+1}=+\infty$.)

In order to see how these pieces fit together,
consider for a moment $t$ and~$z$ to be fixed
and simplify the notation to
\begin{equation}
  \label{eq:piecewise-psi2}
  \psi(x) = A_k e^x + B_k + C_k e^{-x},
  \qquad
  x \in (x_k,x_{k+1}).
\end{equation}
According to \eqref{eq:lax1},
$\psi_{xxx}$ is proportional to a Dirac delta at each point $x_k$,
which means that $\psi$ and $\psi_x$ are continuous
while $\psi_{xx}$ has jump discontinuities:
\begin{equation}
  \label{eq:jump-psi}
  \psi_{xx}(x_k+) - \psi_{xx}(x_k-) = - 2 z \, m_k \psi(x_k)
  \qquad (k=1,\dots,n).
\end{equation}
A straightforward computation translates these continuity and jump
conditions into the following relation between the coefficients in
adjacent intervals:
\begin{equation}
  \label{eq:scattering}
  \begin{pmatrix}
    A_k\\B_k\\C_k
  \end{pmatrix}
  = S_k(z) \,
  \begin{pmatrix}
    A_{k-1}\\B_{k-1}\\C_{k-1}
  \end{pmatrix},
\end{equation}
where the $3\times 3$ matrix $S_k$
(depending on $z$, $x_k$, $m_k$)
is given by
\begin{equation}
  \label{eq:scattering-matrix}
  S_k(z) = I - z m_k
  \begin{pmatrix}
    e^{-x_k} \\ -2  \\ e^{x_k}
  \end{pmatrix}
  \bigl(
    e^{x_k}, 1, e^{-x_k}
  \bigr).
\end{equation}
Here $I$ denotes the $3\times 3$ identity matrix.
Note that $[S_k(z)]^{-1}=S_k(-z)$ and $\det S_k(z) = 1$.

Equation \eqref{eq:scattering} determines $\psi$ completely
once $A_0$, $B_0$, and $C_0$ are known.
The asymptotic condition \eqref{eq:wavefcn-left}
is in the peakon case actually an equality,
$\psi(x)=e^x$ for $x<x_1$,
which corresponds to $A_0=1$ and $B_0=C_0=0$.
So for peakons we get the following special case of
Definition~\ref{def:psi-general}:

\begin{definition}[DP peakon wavefunction]
  \label{def:psi}
  To a given solution
  $\{ x_k(t), m_k(t) \}_{k=1}^n$
  of the DP peakon equations \eqref{eq:peakon-dynamics},
  and to a given complex number $z$,
  we associate the wavefunction
  $\psi(x,t;z)$ given by \eqref{eq:piecewise-psi}
  with coefficients
  \begin{gather}
    \label{eq:A0B0C0}
    \begin{pmatrix}
      A_0(t;z)\\B_0(t;z)\\C_0(t;z)
    \end{pmatrix}
    =
    \begin{pmatrix}
      1\\0\\0
    \end{pmatrix},
    \\
    \label{eq:AkBkCk}
    \begin{pmatrix}
      A_k(t;z)\\B_k(t;z)\\C_k(t;z)
    \end{pmatrix}
    = S_k(z) \, S_{k-1}(z) \cdots S_2(z) \, S_1(z) \,
    \begin{pmatrix}
      1\\0\\0
    \end{pmatrix}
    \qquad (k=1,\dots,n),
  \end{gather}
  where the dependence on $t$ comes from the dependence of
  the matrices $S_k$ on $x_k(t)$ and $m_k(t)$.
\end{definition}

It is clear from \eqref{eq:AkBkCk} and the definition of $S_k$
that $A_k$, $B_k$, $C_k$ are polynomials in $z$ of degree~$k$,
with coefficients depending on all $x_i$'s and $m_i$'s with $i\le k$,
and with constant terms $A_k(t;0)=1$, $B_k(t;0)=C_k(t;0)=0$
(since $S_k(0)=I$).
In fact, it is not hard to see that
\begin{multline}
  \label{eq:AkBkCk-explicit}
  \begin{pmatrix}
    A_k\\B_k\\C_k
  \end{pmatrix}
  =
  \begin{pmatrix}
    1\\0\\0
  \end{pmatrix}
  +
  \\
  \sum_{p=1}^k
  \left[
    \sum_{I\in\binom{[1,k]}{p}}
    \Bigl(
    \prod_{i \in I} m_i
    \Bigr)
    \Bigl(
    \prod_{j=1}^{p-1} (1-e^{x_{i_j}-x_{i_{j+1}}})^2
    \Bigr)
    \begin{pmatrix}
      1 \\ -2 e^{x_{i_p}} \\ e^{2 x_{i_p}}
    \end{pmatrix}
  \right]
  (-z)^p,
\end{multline}
where $\binom{[1,k]}{p}$ is the set of all $p$-element subsets
$I=\left\{ i_1 < \dots < i_p \right\}$ of the integer interval
$[1,k]=\left\{ 1,\dots,k \right\}$.
The empty product ($\prod_{j=1}^{p-1}$ in the case $p=1$)
is interpreted as having the value~$1$.

\begin{example}
  \label{ex:exampleA}
  From \eqref{eq:AkBkCk-explicit} one obtains
  \begin{equation*}
    A_1 = 1 - m_1 z,
    \qquad
    B_1 = 2 m_1 e^{x_1} z,
    \qquad
    C_1 = -m_1 e^{2x_1} z,
  \end{equation*}
  \begin{equation*}
    \begin{split}
      A_2 &= 1 - \bigl[ m_1+m_2 \bigr] z + \bigl[ m_1 m_2 (1-e^{x_1-x_2})^2 \bigr] z^2, \\
      B_2 &= 2 \bigl[ m_1 e^{x_1}+m_2 e^{x_2} \bigr] z - 2 \bigl[ m_1 m_2 (1-e^{x_1-x_2})^2 e^{x_2} \bigr] z^2, \\
      C_2 &= - \bigl[ m_1 e^{2x_1}+m_2 e^{2x_2} \bigr] z + \bigl[ m_1 m_2 (1-e^{x_1-x_2})^2 e^{2x_2} \bigr] z^2,
    \end{split}
  \end{equation*}
  and
  \begin{equation*}
    \begin{split}
      A_3 &= 1 - \bigl[ m_1+m_2+m_3 \bigr] z \\
      &\quad + \bigl[ m_1 m_2 (1-e^{x_1-x_2})^2 + m_1 m_3 (1-e^{x_1-x_3})^2
      + m_2 m_3 (1-e^{x_2-x_3})^2 \bigr] z^2 \\
      &\quad - \bigl[ m_1 m_2 m_3 (1-e^{x_1-x_2})^2 (1-e^{x_2-x_3})^2 \bigr] z^3.
    \end{split}
  \end{equation*}
\end{example}

\begin{remark}
  Although $A_2$, for instance, has the same form for all $n$ considered
  as a function of
  $\{ x_1,x_2,m_1,m_2 \}$,
  the coefficient $A_2(t;z)$ appearing in the rightmost interval
  $x_2 < x < \infty$
  in the two-peakon wavefunction
  is not the same as
  the coefficient $A_2(t;z)$ appearing in the interval $x_2<x<x_3$
  in the three-peakon wavefunction,
  since the functions $\{ x_k(t),m_k(t) \}$ are not the same in the two
  cases.
  In our presentation we always consider $n$ to be fixed (but arbitrary),
  so hopefully no confusion will arise from this
  slight notational ambiguity.
\end{remark}

The rightmost interval $x>x_n$ will be of particular interest,
so we define
\begin{equation}
  \label{eq:ABC}
    \begin{pmatrix}
    A(t;z) \\ B(t;z) \\ C(t;z)
  \end{pmatrix}
  =
  \begin{pmatrix}
    A_n(t;z)\\B_n(t;z)\\C_n(t;z)
  \end{pmatrix}.
\end{equation}

\subsection{Time dependence of $\psi$}
\label{sec:timedep-psi}

Now we take a closer look at the time dependence
of the coefficients $A_k(t;z)$, $B_k(t;z)$, $C_k(t;z)$
appearing in the peakon wavefunction $\psi$
of Definition~\ref{def:psi}.

We know that $\psi(x,t;z)$
must satisfy equation \eqref{eq:lax2}:
\begin{equation*}
  \psi_t = \left[ z^{-1} (1-\partial_x^2) + u_x - u \partial_x \right] \psi,
\end{equation*}
where $u(x,t) = \sum_1^n m_i(t) e^{-\abs{x-x_i(t)}}$.
We claimed before that our choice
\begin{equation*}
  \bigl(A_0(t;z),B_0(t;z),C_0(t;z)\bigr)=(1,0,0)
\end{equation*}
is consistent with \eqref{eq:lax2}.
This is easy to see;
in the leftmost interval $x<x_1(t)$
we have $u(x,t)=\sum_1^n m_i(t) e^{x - x_i(t)}$ ($=u_x$)
and $\psi(x,t;z)=e^x$ ($=\psi_x=\psi_{xx}$),
hence both sides of \eqref{eq:lax2} vanish for $x<x_1(t)$.

Consider now the rightmost interval $x>x_n(t)$,
where we have
$u(x,t)=\sum_1^n m_i(t) e^{x_i(t) - x}$ ($=-u_x$)
and $\psi = A(t;z) \, e^x + B(t;z) + C(t;z) \, e^{-x}$.
Inserting this into \eqref{eq:lax2} yields
\begin{equation*}
  \begin{split}
    A_t e^x + B_t + C_t e^{-x} &=
    \left[ \frac{1-\partial_x^2}{z}
      - e^{-x} \left( \sum_1^n m_i e^{x_i} \right) (1 + \partial_x) \right]
    (A e^x + B + C e^{-x})
    \\
    &= 0\,e^x
    + \left( \frac{B}{z} - 2A \sum_1^n m_i e^{x_i} \right)
    + \left( -B \sum_1^n m_i e^{x_i} \right) e^{-x},
  \end{split}
\end{equation*}
which proves the following:

\begin{proposition}
  \label{prop:timeABC}
  When $\{ x_k, m_k \}_{k=1}^n$
  evolve according to the DP peakon equations \eqref{eq:peakon-dynamics},
  the coefficients in the corresponding peakon wavefunction $\psi(x,t;z)$
  evolve according to the equations
  \begin{equation}
    \label{eq:timeABC}
    A_t = 0,
    \qquad
    B_t = \frac{B}{z} - 2A M_+,
    \qquad
    C_t = -B M_+,
  \end{equation}
  where $M_+ = \sum_{i=1}^n m_i e^{x_i}$.
\end{proposition}

In particular, the polynomial $A(z)$ is independent of~$t$,
so that its coefficients
are constants of motion for \eqref{eq:peakon-dynamics}.
If we write
\begin{equation}
  \label{eq:Acoeff}
  A(z) = 1 - M_1 z + M_2 z - \dots + (-1)^n M_n z^n,
\end{equation}
then equation \eqref{eq:AkBkCk-explicit} gives the explicit formula
\eqref{eq:Mp} for $M_p$ below. It is also useful to write
$A(z)=\prod_{j=1}^n (1-\frac{z}{\lambda _j})$,
where $\{ \lambda_k \}$ by definition are the zeros of $A(z)$
(see Section~\ref{sec:spectral-problem-peakons}), from which we see that
$M_p$ is the $p$th elementary symmetric polynomial in $\{ 1/\lambda_k \}$.

\begin{theorem}
  \label{thm:constants-of-motion}
  The DP $n$-peakon equations \eqref{eq:peakon-dynamics} admit
  $n$ functionally independent constants of motion
  $M_1,\dots,M_n$,
  given explicitly by
  \begin{equation}
    \label{eq:Mp}
    M_p =
    \sum_{I\in\binom{[1,n]}{p}}
    \Bigl(
    \prod_{i \in I} m_i
    \Bigr)
    \Bigl(
    \prod_{j=1}^{p-1} (1-e^{x_{i_j}-x_{i_{j+1}}})^2
    \Bigr),
  \end{equation}
  where $\binom{[1,n]}{p}$ is the set of all $p$-element subsets
  $I=\left\{ i_1 < \dots < i_p \right\}$ of $\left\{ 1,\dots,n \right\}$.
\end{theorem}

\begin{example}
  The constants of motion in the case $n=3$ are
  \begin{equation*}
    \begin{split}
      M_1 &= m_1+m_2+m_3,\\
      M_2 &= m_1 m_2 (1-e^{x_1-x_2})^2 + m_1 m_3 (1-e^{x_1-x_3})^2
      + m_2 m_3 (1-e^{x_2-x_3})^2,\\
      M_3 &= m_1 m_2 m_3 (1-e^{x_1-x_2})^2 (1-e^{x_2-x_3})^2.
    \end{split}
  \end{equation*}
  (Cf.\ the formula for $A_3$ in Example~\ref{ex:exampleA}.)
\end{example}

\subsection{The peakon spectral problem}
\label{sec:spectral-problem-peakons}

In Definition~\ref{def:psi-general} for the wavefunction $\psi$,
we viewed \eqref{eq:lax1} (for fixed $t$ and $z$)
as an initial value problem,
with initial data given by the asymptotic condition at $x=-\infty$.
By imposing constraints both at $x=-\infty$ and at $x=+\infty$,
we turn \eqref{eq:lax1} (for fixed $t$)
into a spectral problem instead:
given $m(x)\ge 0$, find the eigenvalues $z$ such that
\begin{equation}
  \label{eq:spectral-problem-psi}
  \begin{split}
    \psi_x(x) - \psi_{xxx}(x) &= z \, m(x) \, \psi(x),\\
    e^x \, \psi(x) &\to 0 \qquad \text{as $x\to-\infty$},\\
    \psi_x(x) + \psi(x) &\to 0 \qquad \text{as $x\to-\infty$},\\
    e^{-x} \psi(x) &\to 0 \qquad \text{as $x\to+\infty$}
  \end{split}
\end{equation}
has nontrivial solutions $\psi(x)$.
These particular boundary conditions,
which might seem slightly mysterious at this point,
will turn out very natural in the setting of the cubic string;
see Theorem~\ref{thm:liouville}.
In any case, they take a much simpler form in the peakon case,
which is what we are interested in here.

\begin{theorem}
  \label{thm:real-spectrum-peakon}
  Let $m=2\sum_1^n m_i \, \delta_{x_i}$
  be a discrete measure with $m_i>0$ and $x_1<\cdots<x_n$.
  In this case,
  the spectral problem \eqref{eq:spectral-problem-psi}
  has $n$ distinct positive eigenvalues
  \begin{equation*}
    0 < \lambda_1 < \cdots < \lambda_n.
  \end{equation*}
\end{theorem}

\begin{proof}
  We saw in the previous section
  that the solution to \eqref{eq:lax1} in the discrete case
  is given by
  \eqref{eq:piecewise-psi2} and \eqref{eq:scattering},
  and then the boundary conditions in \eqref{eq:spectral-problem-psi}
  reduce to
  \begin{equation*}
    C_0=0, \qquad B_0=0, \qquad A_n=0.
  \end{equation*}
  It is clear that we may normalize $\psi$ by taking $A_0=1$,
  and then $A_n(z)$ is exactly the polynomial we considered before;
  its coefficients depend on the $x_k$'s and $m_k$'s,
  and are given explicitly by \eqref{eq:AkBkCk-explicit} with $k=n$.
  Consequently, the eigenvalues of \eqref{eq:spectral-problem-psi}
  are in the discrete case
  simply the zeros of this $n$th degree polynomial $A(z)=A_n(z)$.

  Now let $\{ x_k,m_k \}$ evolve according to the peakon
  equations \eqref{eq:peakon-dynamics}.
  Even though this changes the measure $m$,
  the eigenvalues of \eqref{eq:spectral-problem-psi} remain unchanged,
  since the coefficients of $A(z)$ are constants of motion
  according to Section~\ref{sec:timedep-psi}.

  As \mbox{$t\to\infty$}, the peakons scatter as described by
  Theorem~\ref{thm:apriori-asymptotics}:
  the $m_k$'s tend to distinct positive values $m_k(\infty)$,
  and $x_{k+1}-x_{k} \to \infty$.
  By Theorem~\ref{thm:constants-of-motion},
  this reduces the coefficients of $A(z)$
  to the elementary symmetric polynomials in $\{ m_k(\infty) \}$,
  so that $A(z)=\prod_{k=1}^n (1-z\,m_k(\infty))$.
  It follows that the eigenvalues are
  $\{ 1/m_k(\infty) \}$,
  hence positive and distinct.
\end{proof}

\begin{remark}
  \label{rem:reverse-order}
  Since $m_1(\infty) < \dots < m_n(\infty)$
  and $m_1(-\infty) > \dots > m_n(-\infty)$ by
  Theorem~\ref{thm:apriori-asymptotics},
  and since the argument of the above proof holds equally well as \mbox{$t\to-\infty$},
  it follows that $m_{n+1-k}(\infty)=1/\lambda_k = m_k(-\infty)$.
  In other words, the same asymptotic velocities occur as
  \mbox{$t\to+\infty$} and \mbox{$t\to-\infty$}, but in reverse order.
  We will study asymptotic peakon properties of this kind in more detail
  in Section~\ref{sec:asymptotics} after giving the
  explicit formulas for the $n$-peakon solution.
\end{remark}

It is clear from the isospectral deformation argument
in the proof above that the numbers $\{ x_k, m_k \}_{k=1}^n$
are not uniquely determined by the eigenvalues $\lambda_k$ alone.
The eigenvalues must be supplemented by further spectral data
in order for the inverse spectral problem of reconstructing $m(x)$
to have a unique solution.

\begin{definition}[Extended spectral data $\{ \lambda_k,b_k,c_k \}$]
  \label{def:spectral-data}
  Given a point in the peakon sector $\mathcal{P}$,
  \begin{equation*}
    x_1<\cdots<x_n
    \qquad\text{and}\qquad
    m_1,\dots,m_n > 0,
  \end{equation*}
  we assign to it a $3 n$-tuple of numbers $\{\lambda _k, b_k, c_k\}_{k=1}^n$
  via the following construction.  Let the polynomials
  $\bigl(A(z),B(z),C(z)\bigr)=\bigl(A_n(z),B_n(z),C_n(z)\bigr)$
  be given by \eqref{eq:AkBkCk-explicit} with $k=n$.
  Let
  \begin{equation}
    \label{eq:def-lambda}
    0 < \lambda_1 < \dots < \lambda_n
  \end{equation}
  be the zeros of $A(z)$
  (which are positive and distinct by
  Theorem~\ref{thm:real-spectrum-peakon}),
  and let $b_k$ and $c_k$ be the residues in
  the partial fraction decompositions
  \begin{equation}
    \label{eq:def-bk-ck}
      \frac{-B(z)}{2z \, A(z)}
      = \sum_{k=1}^n \frac{b_k}{z-\lambda_k},
      \qquad
      \frac{C(z)-B(z)}{2z \, A(z)}
      = \sum_{k=1}^n \frac{c_k}{z-\lambda_k}.
  \end{equation}
  For future convenience (see Theorem~\ref{thm:parfracWZ}), also let
  \begin{equation}
    \label{eq:lambda0-b0-c0}
    \lambda_0=0,
    \qquad
    b_0=1,
    \qquad
    c_0=\frac{1}{2}.
  \end{equation}
\end{definition}

\begin{theorem}
  \label{thm:spectraldata-evolution}
  When $\{ x_k, m_k \}_{k=1}^n$
  evolve according to the DP peakon equations \eqref{eq:peakon-dynamics},
  the extended spectral data $\{ \lambda_k, b_k, c_k \}_{k=1}^n$
  evolve according to the equations
  \begin{equation}
    \label{eq:residue-evolution}
    \dot{\lambda}_k = 0,
    \qquad
    \dot{b}_k = \frac{b_k}{\lambda_k},
    \qquad
    \dot{c}_k = \dot{b}_k + \sum_{j=1}^n \frac{b_k b_j}{\lambda_j}.
  \end{equation}
\end{theorem}

\begin{proof}
  We have already seen that the eigenvalues $\lambda_k$ are constant.
  Let
  \begin{equation*}
    \omega(z) = -\frac{B(z)}{2z \, A(z)}
    = \sum_{k=1}^n \frac{b_k}{z-\lambda_k}
  \end{equation*}
  and
  \begin{equation*}
    \zeta(z) = \frac{C(z)}{2z \, A(z)}
    = \sum_{k=1}^n \frac{c_k-b_k}{z-\lambda_k}.
  \end{equation*}
  Equation \eqref{eq:AkBkCk-explicit} shows that
  $A(0)=1$ and $B(z)=2 M_+ z + O(z^2)$, so that
  $\omega(0)=-M_+$, where $M_+=\sum_1^n m_k e^{x_k}$.
  The time evolution of $A(z)$, $B(z)$, $C(z)$,
  given by \eqref{eq:timeABC}, now translates into
  \begin{equation*}
    \omega_t(z) = -\frac{B_t(z)}{2z \, A(z)}
    = -\frac{B(z)/z - 2A(z)M_+}{2z \, A(z)}
    = \frac{\omega(z) - \omega(0)}{z}
  \end{equation*}
  and
  \begin{equation*}
    \zeta_t(z) = \frac{C_t(z)}{2z \, A(z)}
    = \frac{-B(z) M_+}{2z \, A(z)}
    = - \omega(z) \omega(0).
  \end{equation*}
  Comparing residues at $z=\lambda_k$ on the right- and left-hand sides of
  these equations one obtains at once
  \eqref{eq:residue-evolution}.
\end{proof}

\begin{corollary}
  \label{cor:ck-from-bk}
  The $c_k$'s are determined by the $\lambda_k$'s and the $b_k$'s
  through the formula
  \begin{equation}
    \label{eq:ck}
    c_k = \lambda_k b_k \sum_{j=0}^n \frac{b_j}{\lambda_j+\lambda_k}
    \qquad (k=1,\dots,n).
  \end{equation}
  (Since $c_0=1/2$ and $b_0=1$ by definition,
  the formula holds trivially also in the case $k=0$,
  if interpreted as being a limit as $\lambda_0\to 0+$.)
\end{corollary}

\begin{proof}
  From the second equation in \eqref{eq:residue-evolution} it is immediate that
  $b_k(t)=b_k(0) e^{t/\lambda_k}$.
  Inserting this into the third equation in \eqref{eq:residue-evolution}
  we get
  \begin{equation*}
    \frac{d}{dt}\bigl( c_k(t)-b_k(t) \bigr)
    = \sum_{j=1}^n \frac{b_k(0)b_j(0)}{\lambda_j}
    \exp\left(\frac{t}{\lambda_k}+\frac{t}{\lambda_j}\right).
  \end{equation*}
  From \eqref{eq:AkBkCk-explicit} and Theorem~\ref{thm:apriori-asymptotics}
  it is clear that $C(z)\to 0$ as \mbox{$t\to-\infty$},
  hence $\zeta(z)$ and its residues $c_k-b_k$ also vanish as \mbox{$t\to-\infty$}.
  Thus, integrating the ODE above from $-\infty$ to $t$ yields
  \begin{equation*}
    c_k(t) - b_k(t) =
    \sum_{j=1}^n \frac{b_k(0)b_j(0)}{\lambda_j \left(\frac{1}{\lambda_k}+\frac{1}{\lambda_j}\right)}
    \exp\left(\frac{t}{\lambda_k}+\frac{t}{\lambda_j}\right)
    = \sum_{j=1}^n \frac{\lambda_k \, b_k(t) b_j(t)}{\lambda_j + \lambda_k}.
  \end{equation*}
  Using the definitions $\lambda_0=0$ and $b_0=1$,
  the $b_k$ on the left-hand side can be incorporated in the sum
  by summing from $j=0$ instead of $j=1$,
  and this gives \eqref{eq:ck}.
\end{proof}

\begin{theorem}
  \label{thm:positive-residues}
  The $b_k$'s and $c_k$'s are positive.
\end{theorem}

\begin{proof}
  First we show that $b_k\ne 0$.
  By \eqref{eq:def-bk-ck}, $b_k = -B(\lambda_k)/2\lambda_k A'(\lambda_k)$,
  so it is enough to show that $B(\lambda_k)\ne 0$.
  Let $\psi_k(x)$ denote the eigenfunction corresponding to the
  eigenvalue $\lambda_k$.
  Since $m$ has compact support,
  $\psi_k$ and all its derivatives vanish as $x\to-\infty$,
  while as $x\to+\infty$ the derivatives vanish and
  $\psi_k(\infty)=B(\lambda _k)$.
  Thus it suffices to show that $\psi_k(\infty)\ne 0$.  Since $\psi_k$
  satisfies $\psi_x-\psi_{xxx}=\lambda_k m \psi$ we readily obtain,
  after the standard steps of multiplying by $\psi$, integrating and
  using the boundary conditions, that
  $\frac{1}{2} \psi_k(\infty)^2 =
  \lambda_k \int_{-\infty}^{\infty} m \psi_k^2 \, dx > 0$.
  This proves that $b_k\ne 0$ for all~$k$.

  To prove $ b_k>0$ we use induction on the number of peakons $n$.
  The case $n=1$ follows from explicit formulas.  We assume the claim
  to be valid for an arbitrary configuration of $n-1$ peakons.  Then
  we use the following deformation argument.

  From the explicit formula \eqref{eq:AkBkCk-explicit} for $A(z)$
  and $B(z)$, we see that by letting $x_1$ tend to $-\infty$
  (keeping the other variables fixed) we can make the coefficients in the
  rational function
  \begin{equation*}
    \omega(z) = -\frac{B(z)}{2z\,A(z)} = \sum_{k=1}^n \frac{b_k}{z-\lambda_k}
  \end{equation*}
  come arbitrarily close to the coefficients in the function
  \begin{equation*}
    \wt{\omega} =
    \frac{-(1-z m_1) \wt{B}(z)}{2z\,(1-z m_1) \,\wt{A}(z)},
  \end{equation*}
  (where $\wt{B}$ and $\wt{A}$ are the polynomials computed for the
  $(n-1)$-peakon configuration consisting of all but the first peakon),
  which has a partial fraction decomposition
  \begin{equation*}
    \frac{0}{z-\kappa_1} + \sum_{k=2}^n \frac{d_k}{z-\kappa_k}
    \qquad
    (\kappa_1:=1/m_1).
  \end{equation*}
  (If $1/m_1$ happens to coincide with one of the eigenvalues
  $\kappa_2,\ldots,\kappa_n$
  of the $(n-1)$-peakon configuration,
  then just change $m_1$ a little
  before $x_1$ is sent to $-\infty$
  so that all $\kappa_k$'s are distinct.)
  The poles and residues $\{ \lambda_k,b_k \}_1^n$
  of the rational function $\omega(z)$
  depend continuously on its coefficients,
  which in turn depend continuously on $\{ x_k,m_k \}_1^n$.
  Since the $b_k$'s are nonzero, it follows that they cannot change sign
  during this deformation,
  which brings the $\lambda_k$'s arbitrarily close to the $\kappa_k$'s,
  and $b_1,\ldots,b_n$ arbitrarily close to $0,d_2,\ldots,d_n$.
  By the induction hypothesis $d_k$'s are all positive.
  This implies that $b_2,\ldots,b_n$
  must have been positive to begin with.

  Moreover, we know from the proof of Theorem~\ref{thm:spectraldata-evolution}
  that the strictly positive function $M_+=\sum _{k=1}^n m_k e^{x_k}$ satisfies
  $M_+ = -\omega (0)=\sum_{k=1}^n b_k/\lambda_k $.
  Letting the peakons evolve according to the DP equation,
  $b_1(t)=b_1(0)e^{t/\lambda_1}$ is the dominant term in this sum
  as \mbox{$t\to\infty$}.
  This leads to a contradiction if $b_1<0$, hence $b_1>0$.

  Finally, since all $\lambda_k$'s and $b_k$'s are positive,
  equation \eqref{eq:ck} shows that all $c_k$'s are positive as well.
\end{proof}

\subsection{The explicit $n$-peakon solution}
\label{sec:explicit}

In the extended spectral data $\{ \lambda_k, b_k, c_k \}_{k=1}^n$
investigated in the previous section,
the quantities $c_k$ play only an auxiliary role (although important),
since they are determined by the $\lambda_k$'s and $b_k$'s.
The primary objects are $\{ \lambda_k, b_k \}_{k=1}^n$,
which we will refer to as \emph{spectral variables}.
Definition~\ref{def:spectral-data} gives the prescription
for going from peakon variables $\{ x_k,m_k \}$
to spectral variables $\{ \lambda_k,b_k \}$.
We can view this as a map
\begin{equation}
  \label{eq:spectral-map}
  \begin{aligned}
    \mathcal{S} \;:\;& \mathcal{P} \to \mathcal{R} \\
    &(x,m) \mapsto (\lambda,b),
  \end{aligned}
\end{equation}
where the ``pure peakon sector'' $\mathcal{P}$ is defined by
\eqref{eq:peakon-sector}, while the region of allowable spectral data is
\begin{equation}
  \label{eq:positive-spectraldata}
  \mathcal{R} =
  \bigl\{ (\lambda,b)\in\R^{2n} \;:\; 0<\lambda_1<\cdots<\lambda_n, \; \text{all $b_i>0$} \bigr\}.
\end{equation}
Theorems~\ref{thm:real-spectrum-peakon} and~\ref{thm:positive-residues}
show that $\mathcal{S}$ does map $\mathcal{P}$ into $\mathcal{R}$.
To justify the terminology ``spectral variables'',
we need to show that $\mathcal{S}$ is injective;
in fact we will show that $\mathcal{S}$ is a bijection of
$\mathcal{P}$ onto $\mathcal{R}$, and find $\mathcal{S}^{-1}$ explicitly.
We will do this by transforming the spectral problem
\eqref{eq:spectral-problem-psi}
to the equivalent cubic string spectral problem \eqref{eq:cubic-spectral}
and studying the inverse spectral problem in that setting.
We will state the main result in Theorem~\ref{thm:inverse-problem-peakon}
in order to explore some of its consequences below,
but the central part of its proof
will have to wait until Section~\ref{sec:solution-invprob}.

Once we know how to invert the change of variables
\eqref{eq:spectral-map},
we can also solve the DP $n$-peakon equations \eqref{eq:peakon-dynamics},
since in new variables they simply take the form of the linear equations
\eqref{eq:residue-evolution}.
Using an evolution operator describing the solution to
\eqref{eq:residue-evolution},
\begin{equation}
  \label{eq:evolution-operator}
  \begin{aligned}
    \Phi_t \;:\;& \mathcal{R} \to \mathcal{R}\\
    & (\lambda_k,b_k) \mapsto (\lambda_k,b_k \, e^{t/\lambda_k}),
  \end{aligned}
\end{equation}
we can write the $n$-peakon solution as
\begin{equation}
  \bigl(x(t),m(t)\bigr) =
  \mathcal{S}^{-1} \circ \Phi_t \circ \mathcal{S} \bigl(x(0),m(0)\bigr).
\end{equation}
However, finding the formulas for $\mathcal{S}^{-1}$
is quite a difficult problem,
which will occupy us for most of the remainder of the paper.
We start by introducing necessary notation.

\begin{definition}[Various notation]
  \label{def:notation-galore}
  For $k\ge 2$, let
  \begin{equation}
    \Delta(x_1,\dots,x_k)=\prod_{i<j}(x_i-x_j),
    \qquad
    \Gamma(x_1,\dots,x_k)=\prod_{i<j}(x_i+x_j).
  \end{equation}
  For $k\ge 0$, we recall that $\binom{[1,n]}{k}$ denotes the set of
$k$-element subsets
  $I=\{ i_1<\dots<i_k \}$ of
  the integer interval $[1,n]=\{ 1,\dots,n \}$.
  For $I\in\binom{[1,n]}{k}$,
  let
  \begin{equation}
    \Delta_I=\Delta(\lambda_{i_1},\dots,\lambda_{i_k}),
    \qquad
    \Gamma_I=\Gamma(\lambda_{i_1},\dots,\lambda_{i_k}),
  \end{equation}
  with the special cases
  $\Delta_{\emptyset}=\Gamma_{\emptyset}=\Delta_{\{i\}}=\Gamma_{\{i\}}=1$.
  For two disjoint ordered sets $I$ and $J$ we will also use occasionally
  \begin{equation}
    \Delta^2_{I,J}=\prod_{i\in I, j\in J}(\lambda_{i}-\lambda_{j})^2,
    \qquad
    \Gamma_{I,J}=\prod_{i\in I, j\in J}(\lambda_{i}+\lambda_{j}),
  \end{equation}
  with the convention that the symbols equal $1$ if one or both sets are
  empty.
  (Note that $\Gamma_{I \cup J} = \Gamma_I \Gamma_J \Gamma_{I,J}$,
  and similarly for $\Delta^2$.)
  Furthermore, let
  \begin{equation*}
    \lambda_I = \prod_{i\in I} \lambda_i,
    \qquad
    b_I = \prod_{i\in I} b_i,
  \end{equation*}
with the proviso $\lambda_{\emptyset}=b_{\emptyset}=1$.
\end{definition}

Using this notation we now define the symmetric functions that
appear in the explicit solution formulas for DP peakons.
\begin{definition}[Symmetric functions]
  \label{def:UVW}
  \begin{equation}
    U_k = \sum_{I\in\binom{[1,n]}{k}}
    \frac{\Delta_{I}^2}{\Gamma_I} b_I,
    \qquad
    V_k = \sum_{I\in\binom{[1,n]}{k}}
    \frac{\Delta_{I}^2}{\Gamma_I} \lambda_I b_I,
  \end{equation}
  and
  \begin{equation}
    W_k =
    \begin{vmatrix}
      U_k & V_{k-1} \\
      U_{k+1} & V_k
    \end{vmatrix}
    = U_k V_k - U_{k+1} V_{k-1}.
  \end{equation}
  (To be explicit, $U_0=V_0=1$, and $U_k=V_k=0$ for $k<0$ or $k>n$.)
\end{definition}

We can evaluate $W_k$ explicitly in terms of $(\lambda,b)$ as follows.
\begin{lemma}
  \label{lem:Wk-explicit}
  \begin{equation}
    \label{eq:Wk-explicit}
    \begin{split}
      W_k
      &= \sum_{I\in\binom{[1,n]}{k}} \frac{\Delta_I^4}{\Gamma_I^2} \lambda_I b_I^2
      \\
      &+
      \sum_{m=1}^k
      \sum_{\substack{I\in\binom{[1,n]}{k-m} \\
          J\in\binom{[1,n]}{2m} \\
          I \cap J = \emptyset}}
      b_I^2 b_J
      \Biggl\{
      2^{m+1}
      \frac{\Delta_I^4 \Delta_{I,J}^2 \lambda_{I \cup J}}{\Gamma_I \, \Gamma_{I \cup J}}
      \\
      & \qquad\qquad\qquad\qquad
      \times
      \Biggl(
      \sum_{\substack{C \cup D = J \\ \abs{C}=\abs{D}=m \\ \min(C)<\min(D)}} \Delta_C^2 \Delta_D^2 \Gamma_C \Gamma_D
      \Biggr)
      \Biggr\}.
    \end{split}
  \end{equation}
\end{lemma}

\begin{proof}
  For brevity, let $\Psi_I=\Delta_I^2/\Gamma_I$
  and $\Psi_{I,J}=\Delta_{I,J}^2/\Gamma_{I,J}$.
  Consider first the contribution from $U_k V_k$,
  which is a sum of terms $(\Psi_A b_A) (\Psi_B \lambda_B b_B)$,
  with $A, \, B \in\binom{[1,n]}{k}$.
  Since $b_A$ and $b_B$ each contain only distinct $b_i$'s,
  we have $b_A b_B = b_I^2 b_J$ where
  $I=A\cap B$ and $J=A \triangle B$ (symmetric difference);
  note that $I\cap J=\emptyset$.
  The terms for which $I=A=B$ and $J=\emptyset$ give rise to the
  first sum in \eqref{eq:Wk-explicit}.
  Otherwise, $\abs{I}=k-m$ and $\abs{J}=2m$ for some $1\le m\le k$,
  and the sets $E=A\setminus I$ and $F=B\setminus I$ partition $J=E\cup F$ into
  two disjoint sets with $\abs{E}=\abs{F}=m$.
  With this notation we have
  $(\Psi_A b_A) (\Psi_B \lambda_B b_B) =
  (\Psi_I^2 \Psi_{I,J} \lambda_I) (\Psi_E \Psi_F \lambda_F) (b_I^2 b_J)$.

  The contribution from $U_{k+1} V_{k-1}$ is similar,
  except that $A\in\binom{[1,n]}{k+1}$ and $B\in\binom{[1,n]}{k-1}$
  so that the case $A=B$ can never happen.
  In the disjoint partition $J=G\cup H$, where $G=A\setminus I$ and
  $H=B\setminus I$, we have $\abs{G}=m+1$ and $\abs{H}=m-1$.
  Hence,
  \begin{equation*}
    \begin{split}
      W_k
      &= \sum_{I\in\binom{[1,n]}{k}} \Psi_I^2 \lambda_I b_I^2
      \\
      &+
      \sum_{m=1}^k
      \sum_{\substack{I\in\binom{[1,n]}{k-m} \\
          J\in\binom{[1,n]}{2m} \\
          I \cap J = \emptyset}}
      b_I^2 b_J
      \, \Psi_I^2 \Psi_{I,J} \lambda_I
      \Biggl(
      \sum_{\substack{E \cup F = J \\ \abs{E}=\abs{F}=m}}
      \Psi_{E} \, \Psi_{F} \, \lambda_{F}
      -\sum_{\substack{G \cup H = J \\ \abs{G}=m+1 \\ \abs{H}=m-1}}
      \Psi_{G} \, \Psi_{H} \, \lambda_{H}
      \Biggr).
    \end{split}
  \end{equation*}
  The expression in brackets equals $1/\Gamma_J$
  times the left-hand side of the symmetric polynomial identity
  \begin{multline}
    \label{eq:symm-identity}
    \sum_{\substack{E \cup F = J \\ \abs{E}=\abs{F}=m}}
    \Delta^2_{E} \, \Delta^2_{F} \Gamma_{E,F}\, \lambda_{F}
    -\sum_{\substack{G \cup H = J \\ \abs{G}=m+1 \\ \abs{H}=m-1}}
    \Delta^2_{G} \, \Delta^2_{H} \Gamma_{G,H} \, \lambda_{H}
    \\
    =
    2^{m+1}\,\lambda_J
    \Biggl(
    \sum_{\substack{C \cup D = J \\ \abs{C}=\abs{D}=m \\ \min(C)<\min(D)}} (\Delta_C \Delta_D)^2 \Gamma_C \Gamma_D
    \Biggr),
  \end{multline}
  substitution of which yields the desired formula \eqref{eq:Wk-explicit}.

  To prove \eqref{eq:symm-identity} we use induction on $m=\abs{J}/2$.
  Without loss of generality, we can take $J=\{1,2,\ldots,2m\}$.
  When $m=1$, both sides reduce to $4\,\lambda_1 \lambda_2$.
  For $m>1$, evaluating \eqref{eq:symm-identity} at
  $\lambda_{2m-1}=\lambda_{2m}=c$ yields
  $4\,c^2 \bigl( \prod_{j=1}^{2m-2} (\lambda_j-c)^2 (\lambda_j+c) \bigr)$
  times the left- and right-hand sides, respectively,
  of the $m-1$ case of \eqref{eq:symm-identity},
  which holds by the induction hypothesis.
  Hence, by symmetry, \eqref{eq:symm-identity} holds whenever two
  variables are equal.
  This means that the difference between the left- and the right-hand sides
  is divisible by $\Delta_J$, but since that difference is
  a symmetric polynomial it must be divisible by $\Delta_J^2$.
  However, the difference is of degree $3m-1$ in $\lambda_1$,
  while $\Delta_J^2$ has degree $2(2m-1)$ in $\lambda_1$.
  Since $3m-1 <4m-2$ if $m>1$, we conclude that the difference
  is identically zero,
  and \eqref{eq:symm-identity} is proved.
\end{proof}

\begin{remark}
  Similar polynomial identities will be used later
  in the proof of Lemma~\ref{lem:super-Heine};
  see also Appendix~\ref{sec:app-identities}.
  The constraint $\min(C)<\min(D)$ eliminates redundancy;
  if it is removed, then every term appears twice in the sum,
  and $2^{m+1}$ should be changed to $2^m$.
\end{remark}

\begin{theorem}[Inverse of $\mathcal{S}$]
  \label{thm:inverse-problem-peakon}
  The change of variables $\mathcal{S}: \mathcal{P} \to \mathcal{R}$
  is a bijection. Its inverse
  \begin{equation}
    \label{eq:inverse-spectral-map}
    \begin{aligned}
      \mathcal{S}^{-1} \;:\;& \mathcal{R} \to \mathcal{P} \\
      & (\lambda,b) \mapsto (x,m)
    \end{aligned}
  \end{equation}
  is given by
  \begin{equation}
    \label{eq:n-peakon-solution}
    x_{k'} = \log \frac{U_k}{V_{k-1}},
    \qquad
    m_{k'} =
    \frac{(U_k)^2 \, (V_{k-1})^2}{W_k W_{k-1}}
    \qquad
    (k=1,\dots,n),
  \end{equation}
  where $k'=n+1-k$.
\end{theorem}

\begin{proof}
  For the proof,
  denote the map \eqref{eq:n-peakon-solution} by $\mathcal{T}$
  instead of $\mathcal{S}^{-1}$.
  In Section~\ref{sec:solution-invprob} we will prove that
  $\mathcal{T} \circ \mathcal{S} = \id_{\mathcal{P}}$,
  using the solution of the inverse spectral problem for the
  discrete cubic string.
  In words: provided that $(\lambda,b)=\mathcal{S}(x,m)$ is in the
  range of $\mathcal{S}$, we know that $(x,m)=\mathcal{T}(\lambda,b)$.
  To complete the proof, we need to show that the range of
  $\mathcal{S}$ is all of $\mathcal{R}$
  (that is, all allowable spectral data really correspond
  to a peakon configuration).

  First note that $\mathcal{T}$ maps
  $\mathcal{R}$ into the pure peakon sector $\mathcal{P}$.
  Indeed,
  from the definitions of $U_k$ and $V_k$,
  and from the explicit formula \eqref{eq:Wk-explicit} for $W_k$,
  it is clear that all $U_k$, $V_k$, and $W_k$
  (for $1\le k \le n$)
  are positive when $(\lambda,b)\in\mathcal{R}$;
  then \eqref{eq:n-peakon-solution}
  produces a point $(x,m)=\mathcal{T}(\lambda,b)$ with all $m_i>0$ (obviously),
  and also with $x_1<\cdots<x_n$ since
  \begin{equation*}
    W_k>0
    \quad\Longleftrightarrow\quad
    \frac{U_{k+1}}{V_k} < \frac{U_k}{V_{k-1}}
    \quad\Longleftrightarrow\quad
    x_{k'-1} < x_{k'}.
  \end{equation*}

  Recall the definition of $\mathcal{S}$:
  given $(x,m)\in\mathcal{P}$ we form the polynomials
  $A(z)$ and $B(z)$, whose coefficients are Laurent polynomials in
  $\{e^{x_k},m_k\}_{k=1}^n$.
  Write $A(z;(x,m))$ and $B(z;(x,m))$ to denote this dependence explicitly.
  The $\lambda_k$'s are defined as the zeros of $A(z)$,
  and the $b_k$'s as the residues in $-B(z)/2z\,A(z)$.

  Now we use the algebraic nature of the mappings involved.
  The coefficients in the polynomial $A(z;\mathcal{T}(\lambda,b))$
  are rational functions of $\{\lambda_k,b_k\}_{k=1}^n$
  which, since
  $\mathcal{T} \circ \mathcal{S} = \id_{\mathcal{P}}$,
  agree with the coefficients in $\prod_{k=1}^n (1-z/\lambda_k)$
  on the range of $\mathcal{S}$
  (which is an open set in $\mathcal{R}$ since $\mathcal{S}$ is
  a homeomorphism).
  Hence $A(z;\mathcal{T}(\lambda,b)) = \prod_{k=1}^n (1-z/\lambda_k)$
  identically as a rational function of $(\lambda,b)$;
  in particular this relation holds for all points
  $(\lambda,b)\in\mathcal{R}$.
  Similarly, the coefficients of $B(z;\mathcal{T}(\lambda,b))$
  agree with
  $-2z \bigl(\sum_{i=1}^n b_i/(z-\lambda_i)\bigr) \prod_{k=1}^n (1-z/\lambda_k)$
  on the range of $\mathcal{S}$,
  hence on all of $\mathcal{R}$ since they are rational functions.
  This proves that $\mathcal{S}\circ\mathcal{T}=\id_{\mathcal{R}}$.
  Consequently, the range of $\mathcal{S}$ is $\mathcal{R}$,
  and $\mathcal{T}=\mathcal{S}^{-1}$, as claimed.
\end{proof}

\begin{corollary}[The $n$-peakon solution]
  \label{cor:n-peakon-solution}
  The solution $\{ x_k(t),m_k(t) \}_{k=1}^n$
  of the Degasperis--Procesi $n$-peakon equations
  \eqref{eq:peakon-dynamics}
  is given by \eqref{eq:n-peakon-solution} with
  \begin{equation*}
    b_i(t)=b_i(0) e^{t/\lambda_i}
    \qquad
    (i=1,\dots,n),
  \end{equation*}
  where the constants $\{ \lambda_i, b_i(0) \}_{i=1}^n$
  are determined from the initial conditions
  $\{ x_k(0), m_k(0) \}_{k=1}^n$.
\end{corollary}

\begin{example}[The three-peakon solution]
  \label{ex:threepeakon}
  For $n=1$ and $n=2$,
  equation \eqref{eq:n-peakon-solution} reduces to
  \eqref{eq:onepeakon} and \eqref{eq:twopeakon},
  respectively.
  The solution for $n=3$,
  when simplified by taking into account that some of the
  $U_k$'s and $V_k$'s equal $0$ or $1$,
  takes the form
  \begin{equation}
    \label{eq:threepeakon}
    \begin{aligned}
      x_3(t) &= \log U_1,
      &
      x_2(t) &= \log\frac{U_2}{V_1},
      &
      x_1(t) &= \log\frac{U_3}{V_2},
      \\
      m_3(t) &= \frac{(U_1)^2}{W_1},
      &
      m_2(t) &= \frac{(U_2)^2 (V_1)^2}{W_2 W_1},
      &
      m_1(t) &= \frac{U_3 (V_2)^2}{V_3 W_2},
    \end{aligned}
  \end{equation}
  where, with $b_i=b_i(0) e^{t/\lambda_i}$,
  \begin{equation*}
    \begin{split}
      U_1 &= b_1+b_2+b_3,
      \\
      V_1 &=
      \lambda_1 b_1+\lambda_2 b_2+\lambda_3 b_3,
      \\
      U_2 &=
      \frac{(\lambda_1-\lambda_2)^2}{\lambda_1+\lambda_2} b_1 b_2
      +\frac{(\lambda_1-\lambda_3)^2}{\lambda_1+\lambda_3} b_1 b_3
      +\frac{(\lambda_2-\lambda_3)^2}{\lambda_2+\lambda_3} b_2 b_3,
      \\
      V_2 &=
      \frac{(\lambda_1-\lambda_2)^2}{\lambda_1+\lambda_2} \lambda_1 \lambda_2 b_1 b_2
      +\frac{(\lambda_1-\lambda_3)^2}{\lambda_1+\lambda_3} \lambda_1 \lambda_3 b_1 b_3
      +\frac{(\lambda_2-\lambda_3)^2}{\lambda_2+\lambda_3} \lambda_2 \lambda_3 b_2 b_3,
      \\
      U_3 &=
      \frac{(\lambda_1-\lambda_2)^2
        (\lambda_1-\lambda_3)^2
        (\lambda_2-\lambda_3)^2}
      {(\lambda_1+\lambda_2)
        (\lambda_1+\lambda_3)
        (\lambda_2+\lambda_3)}
      b_1 b_2 b_3,
      \\
      V_3 &=
      \frac{(\lambda_1-\lambda_2)^2
        (\lambda_1-\lambda_3)^2
        (\lambda_2-\lambda_3)^2}
      {(\lambda_1+\lambda_2)
        (\lambda_1+\lambda_3)
        (\lambda_2+\lambda_3)}
      \lambda_1 \lambda_2 \lambda_3
      b_1 b_2 b_3,
    \end{split}
  \end{equation*}
  and consequently
  \begin{equation*}
    \begin{split}
      W_1 &= U_1 V_1 - U_2
      \\
      &=
      \lambda_1 b_1^2 + \lambda_2 b_2^2 + \lambda_3 b_3^2 +
      \frac{4 \lambda_1 \lambda_2}{\lambda_1+\lambda_2} b_1 b_2 +
      \frac{4 \lambda_1 \lambda_3}{\lambda_1+\lambda_3} b_1 b_3 +
      \frac{4 \lambda_2 \lambda_3}{\lambda_2+\lambda_3} b_2 b_3,
      \\
      W_2 &= U_2 V_2 - U_3 V_1
      \\
      &=
      \frac{(\lambda_1-\lambda_2)^4}{(\lambda_1+\lambda_2)^2} \lambda_1 \lambda_2 (b_1 b_2)^2
      + \frac{(\lambda_1-\lambda_3)^4}{(\lambda_1+\lambda_3)^2} \lambda_1 \lambda_3 (b_1 b_3)^2
      + \frac{(\lambda_2-\lambda_3)^4}{(\lambda_2+\lambda_3)^2} \lambda_2 \lambda_3 (b_2 b_3)^2
      \\
      &+
      \frac{4 \lambda_1 \lambda_2 \lambda_3 b_1 b_2 b_3}
      {(\lambda_1+\lambda_2)(\lambda_1+\lambda_3)(\lambda_2+\lambda_3)}
      \times
      \biggl(
        (\lambda_1-\lambda_2)^2 (\lambda_1-\lambda_3)^2 b_1
      \\
      &\qquad\qquad
        + (\lambda_2-\lambda_1)^2 (\lambda_2-\lambda_3)^2 b_2
        + (\lambda_3-\lambda_1)^2 (\lambda_3-\lambda_2)^2 b_3
      \biggr)
      .
    \end{split}
  \end{equation*}
  We take the opportunity here to correct a mistake in our previous paper
  \cite{ls-invprob},
  where the denominator $W_1$ in $m_3(t)$
  was incorrectly stated with $(\lambda_i+\lambda_j)^2$
  instead of $\lambda_i+\lambda_j$.
\end{example}

\subsection{Peakon asymptotics and scattering}
\label{sec:asymptotics}

We will now extract information from the explicit solution formulas
\eqref{eq:n-peakon-solution}
about how the peakons interact, and about their asymptotic behavior as
\mbox{$t\to\pm\infty$}.
Recall that we number the eigenvalues so that
$0 < \lambda_1 < \dots < \lambda_n$,
and consequently
\begin{equation*}
  0 < \frac{1}{\lambda_n} < \dots < \frac{1}{\lambda_1}.
\end{equation*}
Also recall we use $k'=n+1-k$ to denote the complementary index to $k$.

\begin{theorem}[Asymptotics]
  \begin{equation}
    \label{eq:asymptotics-x}
    \begin{aligned}
      \displaystyle
      x_k(t) &\sim
      \frac{t}{\lambda_k} + \log b_k(0) +
      \sum_{i=k+1}^n
      \log \frac{(\lambda_i-\lambda_k)^2}{(\lambda_i+\lambda_k) \lambda_i},
      && \text{as $t\to-\infty$},
      \\
      x_{k'}(t) &\sim
      \displaystyle
      \frac{t}{\lambda_k} + \log b_k(0) +
      \sum_{i=1}^{k-1}
      \log \frac{(\lambda_i-\lambda_k)^2}{(\lambda_i+\lambda_k) \lambda_i},
      && \text{as $t\to+\infty$},
    \end{aligned}
  \end{equation}
  and
  \begin{equation}
    \label{eq:asymptotics-m}
    \begin{aligned}
      m_k(t) &=
      1/\lambda_k
      + O\bigl(e^{t\delta_k}\bigr),
      && \text{as $t\to-\infty$},
      \\
      m_{k'}(t) &=
      1/\lambda_k
      + O\bigl(e^{-t\delta_k}\bigr),
      && \text{as $t\to+\infty$},
    \end{aligned}
  \end{equation}
  where
  \begin{equation}
    \delta_k=
    \begin{cases}
      \frac{1}{\lambda_1}-\frac{1}{\lambda_2}, & k=1,\\
      \min \bigl(
      \frac{1}{\lambda_{k-1}}-\frac{1}{\lambda_{k}},
      \frac{1}{\lambda_k}-\frac{1}{\lambda_{k+1}}
      \bigr),
      & k = 2, \ldots, n-1,\\
      \frac{1}{\lambda_{n-1}}-\frac{1}{\lambda_{n}}, & k=n.
    \end{cases}
  \end{equation}
\end{theorem}

\begin{proof}
  This is simply a matter of identifying the dominant terms among
  the $b_k$'s.
  Because of the ordering of the eigenvalues,
  $b_1(t)=b_1(0) e^{t/\lambda_1}$ dominates as \mbox{$t\to+\infty$}
  and
  $b_n(t)=b_n(0) e^{t/\lambda_n}$ dominates as \mbox{$t\to-\infty$}.
  More generally, $b_1 b_2 \cdots b_k$ grows exponentially
  faster as \mbox{$t\to+\infty$}
  than any other product $b_{i_1} b_{i_2} \cdots b_{i_k}$
  with $i_1 < \cdots < i_k$.
  It follows that, as \mbox{$t\to+\infty$},
  \begin{equation*}
    U_k \sim \frac{(\Delta_{12\dots k})^2}{\Gamma_{12\dots k}}
    b_1 b_2 \cdots b_k,
  \end{equation*}
  and
  \begin{equation*}
    V_k \sim \frac{(\Delta_{12\dots k})^2}{\Gamma_{12\dots k}}
    \lambda_1 \lambda_2 \cdots \lambda_k \,
    b_1 b_2 \cdots b_k,
  \end{equation*}
  while (by similar reasoning)
  for \mbox{$t\to-\infty$} each index on the right-hand side is
  replaced by its complementary value $i'=n+1-i$.
  Hence, as \mbox{$t\to+\infty$},
  \begin{equation*}
    \frac{U_k}{V_{k-1}}
    \sim
    \left(
      \frac{\Delta_{12\dots k}}{\Delta_{12\dots (k-1)}}
    \right)^2
    \frac{\Gamma_{12\dots (k-1)}}{\Gamma_{12\dots k}}
    \frac{b_k}{\lambda_1 \lambda_2 \dots \lambda_{k-1}}
    = b_k(0) e^{t/\lambda_k}
    \prod_{i=1}^{k-1} \frac{(\Delta_{ik})^2}{\Gamma_{ik} \lambda_i}.
  \end{equation*}
  Since $U_k/V_{k-1}=\exp(x_{k'})$,
  the case \mbox{$t\to+\infty$} of \eqref{eq:asymptotics-x}
  follows by taking logarithms.
  The case \mbox{$t\to-\infty$} is similar.

  As for \eqref{eq:asymptotics-m},
  we have already seen (Remark~\ref{rem:reverse-order})
  that
  $m_k(-\infty)=1/\lambda_k=m_{k'}(+\infty)$,
  and this can of course also be concluded by comparing coefficients
  of the dominant term
  $(b_1 \cdots b_{k-1})^4 (b_k)^2$
  (as \mbox{$t\to+\infty$})
  in the numerator and the denominator of the formula for $m_{k'}(t)$.
  As for the correction term,
  for $1<k<n$ a routine computation gives
  \begin{equation*}
    m_{k'}(t)
    = \frac{1}{\lambda_k}
    \frac{(1+A_k\frac{b_{k+1}}{b_k})^2 (1+B_k \frac{b_k}{b_{k-1}})^2}
    {(1+C_k\frac{b_{k+1}}{b_k}) (1+D_{k }\frac{b_k}{b_{k-1}})}
    +\text{lower-order terms}
  \end{equation*}
  as \mbox{$t\to+\infty$},
  where $A_k, B_k, C_k, D_k$
  are some $t$-independent and computable coefficients.
  This implies
  \begin{equation*}
    m_{k'}(t)-\frac{1}{\lambda_k}
    =\alpha_k\frac{b_{k+1}}{b_k} +\beta_k
    \frac{b_k}{b_{k-1}}
    +\text{lower-order terms}
  \end{equation*}
  which gives the stated result as \mbox{$t\to+\infty$}.
  A similar analysis applies to the cases $k=1$ and $k=n$,
  where there are no terms involving $b_{k-1}$ or $b_{k+1}$, respectively.
  Again, the case \mbox{$t\to-\infty$} is similar.
\end{proof}

\begin{corollary}[Phase shifts]
  The peakon with asymptotic velocity $1/\lambda_k$
  as \mbox{$t\to\pm\infty$} experiences the phase shift
  \begin{multline}
    \label{eq:phaseshift}
    \lim_{t\to\infty} \left( x_{k'}(t)-\frac{t}{\lambda_k} \right)
    -\lim_{t\to-\infty} \left( x_k(t)-\frac{t}{\lambda_k} \right)
    =\\
    =\sum_{i=1}^{k-1}
    \log \frac{(\lambda_i-\lambda_k)^2}{(\lambda_i+\lambda_k) \lambda_i}-
\sum_{i=k+1}^{n}\log \frac{(\lambda_i-\lambda_k)^2}{(\lambda_i+\lambda_k) \lambda_i}.
  \end{multline}
\end{corollary}

\begin{remark}
  For every $t$
  there is a well-defined notion of what ``the peakon at site $k$''
  means,
  namely, the peaked wave $m_k(t) \exp(-\abs{x-x_k(t)})$
  described by $m_k$ and $x_k$.
  For other types of soliton equations,
  individual solitons are usually only identifiable as they
  scatter when \mbox{$t\to\pm\infty$}.
  In that case,
  the highest/fastest soliton,
  which is to the far left as \mbox{$t\to-\infty$},
  reemerges to the far right as \mbox{$t\to+\infty$} after some
  interactions during which the identities of the solitons are blurred,
  and similarly for the other solitons.
  According to this point of view,
  the ``$k$th fastest/highest peakon''
  (the one with asymptotic velocity $1/\lambda_k$)
  moves from occupying site $k$ as \mbox{$t\to-\infty$}
  to occupying site $k'$ as \mbox{$t\to+\infty$},
  and does not really have a well-defined identity during the interactions
  causing this change.
  It is the phase shift of this $k$th fastest peakon that is given in
  \eqref{eq:phaseshift}.
\end{remark}

\begin{remark}
  Note that the total phase shift in \eqref{eq:phaseshift}
  equals the sum of phase shifts resulting from pairwise interactions;
  since the peakons reverse their order
  (according to the interpretation above),
  the $k$th fastest peakon is overtaken by the ones initially to its left
  (giving the contribution $\sum_{i=1}^{k-1}$)
  and overtakes the ones initially to its right
  (which contributes $-\sum_{i=k+1}^n$).

  We see that the interaction of a faster peakon overtaking a slower one
  (with asymptotic speeds $c_f=1/\lambda_f$ and $c_s=1/\lambda_s$,
  respectively,
  where $f<s$ so that $c_f>c_s$)
  results in the slower peakon shifting by the amount
  \begin{equation*}
    \log \frac{(\lambda_f-\lambda_s)^2}{(\lambda_f+\lambda_s) \lambda_f}=
    \log \frac{(c_f-c_s)^2}{(c_f+c_s) c_s},
  \end{equation*}
  which is positive iff $c_f > 3 \, c_s$,
  while the shift of the faster peakon is
  \begin{equation*}
    -\log \frac{(\lambda_s-\lambda_f)^2}{(\lambda_s+\lambda_f) \lambda_s}=
    \log \frac{(c_f+c_s) c_f}{(c_f-c_s)^2},
  \end{equation*}
  which is always positive.
  This generalizes the results obtained in \cite{dhh1}
  for the case $n=2$.
\end{remark}

\begin{remark} The phase shifts obtained in the Camassa--Holm peakon
case in \cite{bss-moment}, as well as finite Toda phase shifts obtained
by Moser in \cite{M}, have a slightly different structure;
namely, in the case of CH peakons we have (after some rescaling)
\begin{equation*}
  \sum_{i=1}^{k-1} \log \frac{(\lambda_i-\lambda_k)^2}{\lambda_i^2}-
  \sum_{i=k+1}^{n} \log \frac{(\lambda_i-\lambda_k)^2}{\lambda_i^2}.
\end{equation*}

\end{remark}

\subsection{A few comments about antipeakons}
\label{sec:antipeakons}

So far we have only been discussing the pure peakon case
where all $m_i>0$.
Everything we have said applies (\emph{mutatis mutandis})
also to the pure antipeakon case where all $m_i<0$,
because of the following natural symmetry of the problem.

\begin{proposition}[Antipeakons]
  \label{p-ap invariance}
  Let $\{ x_k(t), m_k(t) \}$ be a global solution to
  \eqref{eq:peakon-dynamics} satisfying
  Assumption~\ref{ass:ordering}
  (pure peakons and ordering).
  Then
  \begin{equation}
    m_k^*(t)=-m_{n+1-k}(-t), \qquad x_k^*(t)=x_{n+1-k}(-t)
  \end{equation}
  is a global solution to \eqref{eq:peakon-dynamics} satisfying
  $m_k^*(t) < 0$,
  $x_1^*(t) > x_2^*(t) > \cdots > x_n^*(t)$.
\end{proposition}

\begin{proof}
  Straightforward.
\end{proof}

However, when both peakons and antipeakons are present,
the situation is more complicated,
because of the singularities that occur after finite time when
a peakon travelling to the right collides with an antipeakon
travelling to the left.
Most of our proofs break down, since they rely on the existence of a
global solution
(Proposition~\ref{prop:global-solution})
and its scattering properties as \mbox{$t\to\pm\infty$}
(Theorem~\ref{thm:apriori-asymptotics}).

We suspect that the spectrum is still real
and simple in the presence of antipeakons,
with the same number of positive (negative) eigenvalues as peakons
(antipeakons).
This is easy to prove for $n=1$ and~$2$, and also agrees with the analogous
situation for the CH peakons \cite{bss-moment}.
However,
in contrast to the CH case,
the spectral measure $\mu=\sum_1^n b_k \, \delta_{\lambda_k}$
need \emph{not} be positive in the mixed peakon-antipeakon case
(it is not hard to find counterexamples for $n=2$).

The explicit solution formulas \eqref{eq:n-peakon-solution}
remain valid as long as they make sense,
because of their algebraic nature.
However, they break down completely if some $\lambda_i+\lambda_j=0$;
for example, in the case of a totally symmetric peakon-antipeakon collision
(equation (6.13) in \cite{dhh1}).
And even if all $\lambda_i+\lambda_j \ne 0$,
the solution provided by \eqref{eq:n-peakon-solution} will in general
not satisfy the ordering assumption $x_1<\cdots<x_n$, or even be defined,
for all~$t$. This is because $U_k$, $V_k$, and $W_k$ need not remain positive
(even if positive at $t=0$) when some $b_k < 0$.
In intervals $t\in(t_1,t_2)$ where the ordering assumption is violated,
\eqref{eq:n-peakon-solution} gives a solution of
\eqref{eq:peakon-dynamics again} (now with some $d_{ij}=x_i-x_j<0$),
but this will not be a solution of the original peakon equations
\eqref{eq:peakon-dynamics} containing $\abs{d_{ij}}$ instead of $d_{ij}$.

All this is also in contrast to the CH case,
where the solution formulas for $x_k(t)$ and $x_{k+1}(t)$
always automatically satisfy $x_k\le x_{k+1}$ for all $t$,
with equality only at the instant of a peakon-antipeakon collision
where $m_k(t)$ and $m_{k+1}(t)$ diverge to $\pm\infty$.
This provides a continuation of the solution past the collision.
It seems as if continuing DP peakon-antipeakon solutions past collisions
would require some kind of gluing of different solutions from different
time intervals,
but the details are not clear at present.

\section{The cubic string}
\label{sec:cubic-string}

The peakon spectral problem
\eqref{eq:spectral-problem-psi}
is equivalent under a change of variables to what we
have called the \emph{cubic string} problem,
an interesting non-selfadjoint third order generalization of the
well-known problem describing the vibrational modes of a string with
nonhomogeneous mass density.
This fact lies at the core of our investigations of the inverse
spectral problem.

\subsection{Liouville transformation to a finite interval}
\label{sec:liouville}

A Liouville transformation is a change of both dependent and
independent variables in order to bring an ODE to its simplest form.
The Liouville transformation that we will use here was inspired
by a similar transformation \cite{bss-moment} relating the spectral problem
for the Camassa--Holm equation to the string equation
$\phi_{yy}(y)=z\,g(y)\,\phi(y)$
with Dirichlet boundary conditions $\phi(\pm 1)=0$.
This selfadjoint spectral problem,
which we briefly review in Appendix~\ref{sec:app-string},
is fundamental in the inverse scattering approach to
finding multi-peakon solutions of the CH equation
\cite{bss-stieltjes,bss-moment}.
As we will see, for the DP equation it is the non-selfadjoint
``cubic string'' that plays the corresponding role.

To get an idea of how the transformation was found,
consider the DP spectral problem \eqref{eq:spectral-problem-psi}
in the simplest possible case $m(x)=0$,
and leave aside the boundary conditions for a moment.
Then the equation is just $(\partial_x-\partial_x^3) \psi(x) = 0$
and the solution is
\begin{equation}
  \label{eq:psiphiABC}
  \begin{split}
    \psi(x)
    &= Ae^x + B + Ce^{-x}
    \\
    &= \frac{(e^x+1)^2}{2e^x}
    \left[
      \frac{A}{2} \left( \frac{2e^x}{e^x+1} \right)^2 +
      \frac{B}{2} \frac{4e^x}{(e^x+1)^2} +
      \frac{C}{2} \left( \frac{2}{e^x+1} \right)^2
    \right]
    \\
    &= \frac{2}{1-y^2}
    \left[
      A \, \frac{(1+y)^2}{2} +
      B \, \frac{(1+y)(1-y)}{2} +
      C \, \frac{(1-y)^2}{2}
    \right],
  \end{split}
\end{equation}
where $y = \tanh(x/2) = \frac{e^x-1}{e^x+1}$.
This change of variables maps $x\in\R$ to $y\in (-1,1)$,
and the inverse transformation is $x=\log \frac{1+y}{1-y}$.
The expression in brackets, call it $\phi(y)$,
is a quadratic polynomial and thus satisfies $-\partial_y^3 \phi = 0$;
the first order term in the operator $\partial_x - \partial_x^3$
has been removed by the transformation.

In the general case, the following holds.

\begin{theorem}
  \label{thm:liouville}
  Under the change of variables
  \begin{equation}
    \label{eq:liouville}
    y=\tanh\frac{x}{2},
    \qquad
    \psi(x)=\frac{2\,\phi(y)}{1-y^2},
  \end{equation}
  the DP spectral problem \eqref{eq:spectral-problem-psi}
  is equivalent to the cubic string problem
  \begin{equation}
    \label{eq:cubic-spectral}
    \begin{split}
      -\phi_{yyy}(y) &= z \, g(y) \, \phi(y)
      \quad\text{for $y \in (-1,1)$},
      \\
      \phi(-1) = \phi_y(-1) &= 0,
      \\
      \phi(1) &= 0,
    \end{split}
  \end{equation}
  where
  \begin{equation}
    \label{eq:gm}
    \left( \frac{1-y^2}{2} \right)^3 g(y) = m(x).
  \end{equation}
  In the discrete case,
  when $m(x)=2 \sum_1^n m_i \, \delta(x-x_i)$,
  equation \eqref{eq:gm} should be interpreted as
  \begin{equation}
    \label{eq:measure-g}
    g(y) = \sum_{i=1}^n g_i \, \delta(y-y_i),
    \quad\text{where}\quad
    y_i = \tanh \frac{x_i}{2},
    \quad
    g_i = \frac{8m_i}{\bigl( 1-y_i^2 \bigr)^2}.
  \end{equation}
\end{theorem}

\begin{proof}
  The first part is a straightforward computation with the chain rule,
  using $\frac{dx}{dy}=\frac{2}{1-y^2}$.
  In particular, the equivalence of the boundary conditions follows from
  \begin{equation*}
    \begin{split}
      \phi(\pm 1) &= \lim_{x\to\pm\infty} \frac{2 e^x}{(e^x+1)^2} \, \psi(x),\\
      \phi_y(-1) &= \lim_{x\to-\infty} \left(
        \psi_x(x) + \frac{1-e^x}{1+e^x} \, \psi(x)
      \right).
    \end{split}
  \end{equation*}
  For the discrete case, $\delta(x-x_i) dx = \delta(y-y_i) dy$,
  hence
  \begin{equation*}
    \delta(x-x_i) = \frac{\delta(y-y_i)}{\frac{dx}{dy}(y_i)}
  \end{equation*}
  and the statement follows.
\end{proof}

By analogy with the ordinary discrete string, we will refer to the
quantities $g_i$ as point masses at the positions $y_i$, although
physically speaking this terminology is perhaps not justified.
We also define $y_0=-1$ and $y_{n+1}=1$,
which is consistent with our convention that
$x_0=-\infty$ and $x_{n+1}=\infty$.

\begin{remark}
  For comparison, we remark that
  the Liouville transformation $y=\tanh x$, $\psi(x) = \phi(y) / \sqrt{1-y^2}$
  used in the CH case \cite{bss-moment}
  maps $\psi(x)=Ae^x+Be^{-x} \in \ker(\partial_x^2-1)$
  to $\phi(y)=A(1+y)+B(1-y) \in \ker(\partial_y^2)$.
\end{remark}

Since the adjoint of \eqref{eq:cubic-spectral}
involves two boundary conditions
at the right endpoint and one at the left endpoint,
\eqref{eq:cubic-spectral} is not a selfadjoint problem.
Hence there is no \emph{a priori} reason to suspect that the
eigenvalues need to be real.
However, this follows in the discrete case immediately from
Theorems~\ref{thm:liouville}
and~\ref{thm:real-spectrum-peakon}.

\begin{theorem}
  \label{thm:real-spectrum-discrete-cubic}
  The discrete cubic string problem \eqref{eq:cubic-spectral},
  \eqref{eq:measure-g},
  with all $g_i>0$,
  has $n$ distinct positive eigenvalues
  \begin{equation*}
    0 < \lambda_1 < \cdots < \lambda_n.
  \end{equation*}
\end{theorem}

In the general case, when the mass distribution $g(y)$ is not discrete,
there is an infinite sequence of eigenvalues
\begin{equation*}
  0 < \lambda_1 < \lambda_2 < \cdots,
\end{equation*}
which we will prove in Section~\ref{sec:GK} using the beautiful theory
of oscillatory kernels developed by Gantmacher and Krein.

\subsection{The discrete cubic string wavefunction $\phi(y;z)$}
\label{sec:wavefcn-cubic}

Because of Theorem~\ref{thm:liouville},
all notions defined in the context of the peakon spectral problem
\eqref{eq:spectral-problem-psi}
have counterparts in the cubic string setting.
The DP equation induces an isospectral deformation of the cubic string,
but we mostly consider a fixed time~$t$
and suppress the time dependence in the notation.

The DP wavefunction $\psi(x;z)$ of Definition~\ref{def:psi}
is mapped by the Liouville transformation \eqref{eq:liouville} to
the \emph{cubic string wavefunction} $\phi(y;z)$,
which by definition is the solution of the initial value problem
\begin{equation}
  \label{eq:cubicstring}
  \begin{split}
    \phi_{yyy}(y) + z \, g(y) \, \phi(y) &= 0
    \quad\text{for $y \in (-1,1)$},
    \\
    \phi(-1) = \phi_y(-1) &= 0,
    \\
    \phi_{yy}(-1) &= 1.
  \end{split}
\end{equation}
Again, we are mainly interested in the discrete case
\begin{gather*}
  g(y)=\sum_1^n g_i \, \delta(y-y_i),
  \qquad
  g_i > 0,
  \\
  -1 = y_0 < y_1 < \cdots < y_n < y_{n+1}=+1.
\end{gather*}
From \eqref{eq:psiphiABC} it is clear that
the peakon wavefunction, piecewise given by
\begin{equation*}
  \psi(x;z) = A_k(z) e^x + B_k(z) + C_k(z) e^{-x},
  \qquad
  x \in (x_k,x_{k+1})
\end{equation*}
according to \eqref{eq:piecewise-psi2},
corresponds to the \emph{discrete cubic string wavefunction}
\begin{equation}
  \label{eq:piecewise-phi}
  \phi(y;z) = A_k(z) \frac{(1+y)^2}{2} + B_k(z) \frac{1-y^2}{2} + C_k(z) \frac{(1-y)^2}{2},
  \qquad
  y \in (y_k,y_{k+1}),
\end{equation}
which is piecewise a quadratic polynomial in~$y$.
We will denote by
\begin{equation}
  \label{eq:lk}
  l_k = y_{k+1} - y_k
\end{equation}
the length of the $k$th interval $(y_k,y_{k+1})$, for $k=0,\dots,n$.
The coefficients $A_k$, $B_k$, $C_k$ in \eqref{eq:piecewise-phi}
are the same as before and are
given by \eqref{eq:AkBkCk-explicit}.
In particular, $A_0=1$ and $B_0=C_0=0$,
so that $\phi(y;z)=\frac{1}{2} (1+y)^2$ for $-1\le y \le 1$.
However, we will not make much use of these coefficients from now on;
instead we will keep track of $\phi(y;z)$
through the values of the function and its derivatives at the points $y_k$.
We will use the vector notation
\begin{equation*}
  \Phi(y;z) =
  \begin{pmatrix}
    \phi(y;z) \\ \phi_y(y;z) \\ \phi_{yy}(y;z)
  \end{pmatrix},
\end{equation*}
sometimes omitting $z$ to simplify the notation.

By \eqref{eq:cubicstring},
$\phi_{yyy}(y)$ is singular (proportional to the delta function)
at the points $y_1,\dots,y_n$,
and zero elsewhere.
Thus we require $\phi(y)$ and $\phi_y(y)$ to be continuous,
and $\phi_{yy}(y)$ to be piecewise constant with jump discontinuities
at $y_1,\dots,y_n$:
\begin{equation}
  \label{eq:jump-phi}
  \phi_{yy}(y_k+) - \phi_{yy}(y_k-) = -z \, g_k \, \phi(y_k).
\end{equation}
These jump conditions can be expressed as
\begin{equation}
  \label{eq:jump-matrix-phi}
  \Phi(y_k+) = G_k(z) \Phi(y_k-),
  \quad\text{where}\quad
  G_k(z) =
  \begin{pmatrix}
    1 & 0 & 0 \\
    0 & 1 & 0 \\
    -z\,g_k & 0 & 1
  \end{pmatrix}.
\end{equation}
Further,
\begin{equation*}
  \phi(y) = \phi(y_k) + \phi_y(y_k) \, (y-y_k)
  + \phi_{yy}(y_k+) \, \frac{(y-y_k)^2}{2}
\end{equation*}
for $y_k \le y \le y_{k+1}$.
Evaluating this expression and its derivatives
at $y=y_{k+1}$ tells us how $\Phi$ propagates
from one jump to the next:
\begin{equation}
  \label{eq:transfer-matrix-phi}
  \Phi(y_{k+1}-) = L_k \Phi(y_k+),
  \quad\text{where}\quad
  L_k =
  \begin{pmatrix}
    1 & l_k & l_k^2/2 \\
    0 & 1 & l_k \\
    0 & 0 & 1
  \end{pmatrix}.
\end{equation}

Clearly, the vectors $\Phi(y_k\pm;z)$ will consist of polynomials
in $z$ (with coefficients that depend on $g_i$'s and $l_i$'s).
The degrees are as follows, for $k\ge 1$:
as \mbox{$z\to\infty$},
\begin{equation}
  \label{eq:degreePhi}
  \Phi(y_k-;z) = L_{k-1} \, G_{k-1}(z) \dots L_1 \, G_1(z) \, L_0
  \begin{pmatrix}
    0\\0\\1
  \end{pmatrix}
  =
  \begin{pmatrix}
    O(z^{k-1})\\
    O(z^{k-1})\\
    O(z^{k-1})
  \end{pmatrix}
\end{equation}
and
\begin{equation}
  \label{eq:degreePhi2}
  \Phi(y_k+;z) = G_k(z) \, \Phi(y_k-;z)
  =
  \begin{pmatrix}
    O(z^{k-1})\\
    O(z^{k-1})\\
    O(z^k)
  \end{pmatrix}.
\end{equation}
It is immediate that $G_k(z)^{-1} = G_k(-z)$
and $L_k(l_k)^{-1} = L_k(-l_k)$.
Some slightly less obvious relations,
which are crucial in the solution of the inverse spectral problem,
concern the transposed inverses:
\begin{equation}
  \label{eq:conjugateJ}
  (L_k^{-1})^t = J L_k J
  \quad\text{and}\quad
  (G_k(z)^{-1})^t = J G_k(-z) J,
\end{equation}
where
\begin{equation}
  \label{eq:J}
  J =
  \begin{pmatrix}
    0&0&1\\
    0&-1&0\\
    1&0&0
  \end{pmatrix}
  = J^{-1}.
\end{equation}

\subsection{Weyl functions}
\label{sec:weyl}

The eigenvalues of the cubic string problem \eqref{eq:cubic-spectral}
are the zeros of the function $\phi(1;z)$,
where $\phi(y;z)$ is the cubic string wavefunction defined in
the previous section as the solution of the initial value problem
\eqref{eq:cubicstring}.
The extended spectral data are encoded in a pair of \emph{Weyl functions}.

\begin{definition}
  \label{def:weyl}
  The Weyl functions for the cubic string are
  \begin{equation}\label{eq:weyl}
    W(z) = \frac{\phi_y(1;z)}{\phi(1;z)},
    \qquad
    Z(z) = \frac{\phi_{yy}(1;z)}{\phi(1;z)},
  \end{equation}
  where $\phi(y;z)$ is the cubic string wavefunction.
\end{definition}

\begin{theorem}\label{thm:parfracWZ}
  In the discrete case $g(y)=\sum_1^n g_k \, \delta(y-y_k)$, $g_k>0$,
  the Weyl functions admit the partial fraction decompositions
  \begin{align}
    \label{eq:parfracW}
    \frac{W(z)}{z}&=\frac{1}{z}+\sum_{k=1}^n \frac{b_k}{z-\lambda_k}
    =\sum_{k=0}^n \frac{b_k}{z-\lambda_k},
    \\
    \label{eq:parfracZ}
    \frac{Z(z)}{z}&=\frac{1/2}{z}+\sum_{k=1}^n \frac{c_k}{z-\lambda_k}
     =\sum_{k=0}^n \frac{c_k}{z-\lambda_k},
  \end{align}
  where $\{ \lambda_k,b_k,c_k \}_{k=0}^n$
  constitute the extended spectral data of the corresponding peakon
spectral problem
  (see Theorem~\ref{thm:liouville} and Definition~\ref{def:spectral-data}).
\end{theorem}

\begin{proof}
  In the discrete case, $\phi(1;z)$ is an $n$th degree polynomial in~$z$.
  The eigenvalues $\lambda_1,\dots,\lambda_n$,
  which are simple and positive by
  Theorem~\ref{thm:real-spectrum-discrete-cubic},
  are the zeros of $\phi(1;z)$.
  This shows that $W(z)/z$ and $Z(z)/z$ have only simple poles,
  precisely at $z=0=\lambda_0$ and at each eigenvalue $z=\lambda_k$.
  To compute the residues at $z=0$, we use that
  $\phi(y;0)=\frac{1}{2}(1+y)^2$,
  which gives $b_0=\frac{\phi_y(1;0)}{\phi(1;0)}=\frac{2}{2}$
  and $c_0=\frac{\phi_{yy}(1;0)}{\phi(1;0)}=\frac{1}{2}$
  in agreement with our definitions.

  In the rightmost interval $y_n<y<1$ we have, by \eqref{eq:piecewise-phi},
  \begin{equation*}
    \phi(y;z) =
    A(z) \frac{(1+y)^2}{2} + B(z) \frac{1-y^2}{2} + C(z) \frac{(1-y)^2}{2},
  \end{equation*}
  hence
  \begin{equation*}
    \frac{W(z)}{z} = \frac{\phi_y(1;z)}{z\,\phi(1;z)}
    = \frac{2A(z)-B(z)}{2z \, A(z)}
    = \frac{1}{z} - \frac{B(z)}{2z \, A(z)}
  \end{equation*}
  and
  \begin{equation*}
    \frac{Z(z)}{z} = \frac{\phi_{yy}(1;z)}{z\,\phi(1;z)}
    = \frac{A(z)-B(z)+C(z)}{2z \, A(z)}
    = \frac{1/2}{z} + \frac{C(z)-B(z)}{2z \, A(z)}.
  \end{equation*}
  Comparison with \eqref{eq:def-bk-ck} finishes the proof.
\end{proof}

\begin{corollary}
  \label{cor:ZfromW}
  The second Weyl function $Z(z)$ is determined by the
  first Weyl function $W(z)$ through equation \eqref{eq:ck}
  expressing $c_k$ in terms of $\lambda_k$'s and $b_k$'s.
\end{corollary}

This fact, which is absolutely fundamental to the solution of the
inverse spectral problem, is far from obvious from the definition of
the Weyl functions in terms of a given discrete cubic string.
In fact, the proof of \eqref{eq:ck} in
Section~\ref{sec:spectral-problem-peakons}
was only made possible by knowing how the spectral data change
under the peakon evolution.
In other words, in order to prove Corollary~\ref{cor:ZfromW}
we needed to know about the
isospectral deformation of the cubic string
induced by the DP equation.

\subsection{Gantmacher--Krein theory and the cubic string}
\label{sec:GK}

Before we turn to the inverse spectral problem,
we would like to show how the cubic string fits into
the Gantmacher--Krein theory of oscillatory kernels \cite{gantmacher-krein}.
We begin by collecting a few facts about
three classes of matrices with nonnegative entries, see e.g \cite{Fomin} or
\cite{Gekhtman}.

\begin{definition}
  An $n\times n$ matrix $A$ is called
  \emph{totally positive}
  (\emph{totally nonnegative})
  if every minor of $A$ is positive (nonnegative).
  A totally nonnegative matrix $A$ is called \emph{oscillatory}
  if some power of it is totally positive.
\end{definition}

Oscillatory matrices can be thought of as being somewhere between
totally nonnegative and totally positive matrices.
The most pertinent property of these classes of matrices is the following.

\begin{theorem}
  All eigenvalues of a \emph{totally positive} matrix are positive and
  of algebraic multiplicity one.
  All eigenvalues of a \emph{totally nonnegative} matrix are
  nonnegative, but in general of arbitrary multiplicity.
\end{theorem}

\begin{corollary}
  All eigenvalues of an \emph{oscillatory} matrix are positive and of
  algebraic multiplicity one.
\end{corollary}

\begin{definition}
  Suppose $I$ is an open interval of $\R$.
  A (necessarily positive) continuous function
  $K(y,s)$, $y,s \in I$ is called \emph{oscillatory} if for every
  choice of points $y_1 < y_2 < \cdots < y_n \in I$ the matrix
  $[K(y_i,y_j)]_{i,j=1,\dots,n}$ is oscillatory.
\end{definition}

Consider the integral equation
\begin{equation}\label{eq:integral}
  \phi(y) = z \int_I K(y,s) \, \phi(s) \, d\sigma(s),
\end{equation}
where the integral is taken in the sense of Stieltjes,
with $\sigma$ a non-decreasing function on $I$.
In the above, $z$ plays the role of an eigenvalue and
the sought solution $\phi$ is the corresponding eigenfunction.
The central result of the theory developed by Gantmacher and
Krein is the following theorem.
\begin{theorem}[Gantmacher--Krein]\label{thm:GK}
  If the kernel $K$ of the integral equation \eqref{eq:integral}
  is oscillatory, then:
  \begin{enumerate}
  \item The eigenvalues are all positive and simple:
    $0<\lambda_1<\lambda_2<\cdots$.
    If the function $\sigma(s)$ has only a finite number $n$ of points
    of growth, then there are $n$ eigenvalues.
    Otherwise, there are infinitely many eigenvalues.
  \item The eigenfunction $\phi_1$ corresponding to the smallest eigenvalue
    $\lambda_1$ has no zeros in the interval $I$.
  \item For every $j>1$, the eigenfunction $\phi_j$ corresponding
    to the $j$th eigenvalue $\lambda_j$ has exactly $j-1$ nodal points
    in $I$ (i.e., it has $j-1$ zeros and changes its sign at each).
  \end{enumerate}
\end{theorem}

The integral equation \eqref{eq:integral} occurs most naturally in the
context of boundary value problems for ordinary differential
operators.

\begin{theorem}[Krein \cite{krein}]\label{thm:Krein}
  Let
  $L(\phi(y)) = \sum_{k=0}^n l_k(y) \, \phi^{(k)}(y)$,
  where $l_n(y)>0$,
  and consider the boundary value problem
  \begin{gather*}
    L(\phi)=0,\\
    \phi(a)=\phi'(a)=\dots=\phi^{(p-1)}(a)=0,\\
    \phi(b)=\phi'(b)=\dots=\phi^{(q-1)}(b)=0,
  \end{gather*}
  where $p+q=n \ge 2$.
  Let $G(y,s)$ be the Green function corresponding to this
  boundary value problem and let $W(f_1,f_2,\dots,f_k)$ denote the
  Wronskian of the $k$ times differentiable functions
  $f_1,f_2,\dots,f_k$.
  Then $(-1)^q \, G(y,s)$ is oscillatory if and only if
  there exist $p$ solutions
  $\omega_1(y)$, $\omega_2(y)$, \ldots, $\omega_p(y)$
  of the boundary value problem
  \begin{gather*}
    L(\phi)=0, \\
    \phi(b)=\phi'(b)=\dots=\phi^{(q-1)}(b)=0
  \end{gather*}
  and $q$ solutions
  $\omega_{p+1}(y)$, $\omega_{p+2}(y)$, \ldots, $\omega_n(y)$
  of the boundary value problem
  \begin{gather*}
    L(\phi)=0, \\
    \phi(a)=\phi'(a)=\dots=\phi^{(p-1)}(a)=0
  \end{gather*}
  such that, for all $y\in I$,
  \begin{equation*}
    \omega_1 > 0,\quad
    W(\omega_1,\omega_2) > 0,\quad
    \ldots\;,\quad
    W(\omega_1,\omega_2,\ldots,\omega_n) > 0.
  \end{equation*}
\end{theorem}

The cubic string equation \eqref{eq:cubic-spectral} is a special case of this
setup that corresponds to $p=2$, $q=1$, $L=d^3/dy^3$, and $I=(-1,1)$.
The Green function of the cubic string is the solution of
\begin{gather*}
  G_{yyy}(y,s) = \delta(y-s), \\
  G(-1,s) = G_y(-1,s) = 0,
  \qquad
  G(1,s) = 0,
\end{gather*}
namely
\begin{equation}
  \label{eq:cubicGreen}
  G(y,s) =
  \begin{cases}
    \ds -\frac{(s-1)^2 (y+1)^2}{8},
    & -1 \le y \le s \le 1,\\
    \ds \frac{(s+1)^2 (y-1)^2}{8} - \frac{(s^2-1)(y^2-1)}{4},
    & -1 \le s \le y \le 1.
  \end{cases}
\end{equation}

\begin{lemma}
  Let $G(y,s)$ be the Green function \eqref{eq:cubicGreen}
  for the cubic string.
  Then $K(y,s)=-G(y,s)$
  is oscillatory.
\end{lemma}

\begin{proof}
  By direct computation one verifies that
  $\omega_1(y)=1-y$,
  $\omega_2(y)=-(1-y)^2$,
  and $\omega_3(y)=(1+y)^2$
  satisfy the conditions in Theorem~\ref{thm:Krein}.
\end{proof}

The cubic string problem \eqref{eq:cubic-spectral}
is equivalent to the integral equation
\begin{equation}
  \label{eq:cubic-integral}
  \phi(y)=z\int_{-1}^1 K(y,s) \, \phi(s) \, d\sigma(s),
\end{equation}
where $K(y,s)=-G(y,s)$ and $d\sigma(s)=g(s)\, ds$.
In this formulation we can include any kind of (positive) mass distribution,
by letting the non-decreasing function $\sigma(s)$ equal the
accumulated mass in the interval $-1\le y \le s$.
Since $K$ is oscillatory, Theorem~\ref{thm:GK}
gives the following more general version of
Theorem~\ref{thm:real-spectrum-discrete-cubic},
valid not only in the discrete case.

\begin{theorem}
  The spectrum of the cubic string problem \eqref{eq:cubic-integral}
  is positive and simple:
  $0<\lambda_1<\lambda_2<\cdots$.
\end{theorem}

Theorem~\ref{thm:GK} also provides an alternative proof of the
positivity of the $b_k$'s.

\begin{theorem}
  The residues $b_k$ in the Weyl function $W(z)$
  of the discrete cubic string are positive.
  (See equation \eqref{eq:parfracW}.)
\end{theorem}

\begin{proof}
  We have
  \begin{equation}
    \label{eq:ohlala}
    b_k = \res_{z=\lambda_k} \frac{\phi_y(1;z)}{z\,\phi(1;z)}
    = \frac{-\phi_y(1;\lambda_k)}{2\prod_{j\ne k} (1-\lambda_k/\lambda_j)},
  \end{equation}
  where $\phi(y;z)=2\prod_{j=1}^n (1-z/\lambda_k)$
  is the solution to the cubic string initial value
  problem with $\phi(-1;z)=\phi_y(-1;z)=0$ and $\phi_{yy}(1;z)=1$.
  In other words, $\phi(y;\lambda_k)$ is the $k$th eigenfunction
  of the cubic string boundary value problem,
  and as such it has $k-1$ nodal points according to Theorem~\ref{thm:GK}.
  It follows that $(-1)^k \phi_y(1;\lambda_k) \ge 0$.
  Suppose equality holds.
  Then $\phi(y;\lambda_k)$ is simultaneously an eigenfunction of
  our usual cubic string and of the mirrored cubic string ($p=1, q=2$)
  with boundary conditions $\phi(1;z)=\phi_y(1;z)=\phi(-1;z)=0$.
  This is a contradiction,
  since it follows immediately from Theorems~\ref{thm:GK} and \ref{thm:Krein}
  that the mirrored cubic string has negative eigenvalues.
  Hence, $(-1)^k \phi_y(1;\lambda_k) > 0$.
  And since precisely $k-1$ factors in the denominator of
  \eqref{eq:ohlala} are negative,
  it follows that $b_k>0$.
\end{proof}

\section{Inverse problem for the discrete cubic string}
\label{sec:inverse-cubic}

Our aim in this section is to give an explicit solution of the inverse
spectral problem for the discrete cubic string \eqref{eq:cubic-spectral},
\eqref{eq:measure-g}.
This will also provide us with what is needed to complete the proof of
Theorem~\ref{thm:inverse-problem-peakon} regarding the corresponding
inverse spectral problem for DP peakons.

The inverse problem for the ordinary (not cubic) discrete string plays a
fundamental role in solving the peakon problem for the
Camassa--Holm equation
\cite{bss-stieltjes,bss-moment}.
For the reader's convenience, we sketch the Stieltjes--Krein solution
to that inverse problem in Appendix~\ref{sec:app-string}, with emphasis on the
structures common to both inverse problems.

\subsection{General setup}

We begin by stating the problem precisely.

\begin{definition}[Inverse problem]
  \label{def:cubicstring IP}
  Consider a discrete cubic string specified by a positive measure
  $g(y)=\sum_{i=1}^n g_i \, \delta_{y_i} (y)$
  whose support satisfies $-1<y_1<y_2<\dotsc<y_n <1$.
  The inverse spectral problem is to determine the measure $g(y)$
  given the Weyl function $W(z)$ of the cubic string:
  \begin{equation*}
    \frac {W(z)}{z}
    = \sum_{i=0}^n \frac{b_i}{z-\lambda_i}
    =\int \frac{d\mu(\lambda)}{z-\lambda}
  \end{equation*}
  where
  \begin{equation*}
    \mu=\sum_{i=0}^n b_i \, \delta_{\lambda_i},
  \end{equation*}
  $b_0=1$, $b_k>0$, and
  $0=\lambda_0<\lambda_1<\cdots<\lambda_n$,
\end{definition}

Recall that by Definition~\ref{def:weyl} the Weyl functions are
\begin{equation*}
  W(z)=\frac{\phi_y(1;z)}{\phi(1;z)},
  \qquad
  Z(z)=\frac{\phi_{yy}(1;z)}{\phi(1;z)},
\end{equation*}
where, according to \eqref{eq:cubicstring}, \eqref{eq:jump-matrix-phi},
\eqref{eq:transfer-matrix-phi},
\begin{equation*}
  \Phi(1;z)
  =
  \begin{pmatrix}
    \phi(1;z) \\
    \phi_y(1;z) \\
    \phi_{yy}(1;z)
  \end{pmatrix}
  =
  L_n \, G_n(z) \, L_{n-1} \, G_{n-1}(z) \cdots L_1 \, G_1(z) \, L_0
  \begin{pmatrix}
    0 \\ 0 \\ 1
  \end{pmatrix}.
\end{equation*}
Also remember that $Z$ is determined by $W$ (Corollary~\ref{cor:ZfromW}).

The analogy with the case of Stieltjes continued fraction (see Appendix~\ref{sec:app-string})
suggests that the $3\times 3$ matrices
\begin{equation*}
  \begin{split}
  a^{(1)}(z) &= L_n\\
  a^{(2)}(z) &= L_n \, G_n(z)\\
  a^{(3)}(z) &= L_n \, G_n(z) \, L_{n-1}\\
  a^{(4)}(z) &= L_n \, G_n(z) \, L_{n-1} \, G_{n-1}(z)\\
  & \;\;\vdots
  \end{split}
\end{equation*}
might contain convergent-like quantities from which one can construct
successively better rational approximations to $W(z)$ and $Z(z)$.
Likewise, from the vectors
\begin{alignat*}{2}
  \Phi(y_0+;z) &=&  (0,0,1)^t  \\
  \Phi(y_1-;z) &=& L_0 \, (0,0,1)^t \\
  \Phi(y_1+;z) &=& G_1(z) \, L_0 \, (0,0,1)^t \\
  \Phi(y_2-;z) &=& L_1 \, G_1(z) \, L_0 \, (0,0,1)^t\\
  \Phi(y_2+;z) &=&\: G_2(z) \, L_1 \, G_1(z) \, L_0 \, (0,0,1)^t\\
  & \;\;\vdots &
\end{alignat*}
one should be able to form remainder-like quantities.
This is indeed the case, as we shall soon show.

In contrast to the classical Stieltjes theory,
it is not necessary to consider both the even and the odd case.
This is because the matrices $a^{(2k)}(z)$ and $a^{(2k-1)}(z)$
differ only in the first column,
and it turns out that the information that we want to extract can be found
in their common second column.
We choose here to study the even case,
where the starting point is the relation
\begin{equation}
  \label{eq:startingpoint}
  \Phi(1;z) = a^{(2k)}(z) \, \Phi(y_{n-k+1}-;z).
\end{equation}
We will for the most part regard $k\in\{1,\dots,n\}$ as fixed (but arbitrary),
and denote the matrix $a^{(2k)}(z)$ simply by $a(z)$ or~$a$.
We collect some basic facts about it in the following lemma.
As in Section~\ref{sec:explicit}, it is convenient to let $k'=n+1-k$.
\begin{lemma}
  \label{lem:a-facts}
  The matrix
  \begin{equation}
    \label{eq:defevena}
    a(z) = a^{(2k)}(z) =
    L_n \, G_n(z) \, L_{n-1} \, G_{n-1}(z) \cdots L_{k'} \, G_{k'}(z)
  \end{equation}
  satisfies $\det a(z) = 1$.
  The entries in its first column are polynomials in~$z$ of degree $k$,
  while the remaining entries are polynomials of degree $k-1$.
  In the second column we note in particular that
  \begin{equation}
    \label{eq:a12}
    a_{12}(z) =
    l_{k'} \left( \prod_{i=k'+1}^n \frac{-g_i l_i^2}{2} \right) \, z^{k-1}
    + \ldots +
    \left( \sum_{i=k'}^n l_i \right)
  \end{equation}
  and
  \begin{equation}
    \label{eq:a22-a32-at-zero}
    a_{22}(0) = 1,
    \qquad
    a_{32}(0) = 0.
  \end{equation}
  All $2\times 2$ minors of $a(z)$ are polynomials of degree at most $k$.
  \end{lemma}

\begin{proof}
  The matrices $G_i(z)$ and $L_i$ have determinant one,
  hence so does $a(z)$.
  Since $G_i(z)$ is linear in $z$,
  it is clear that $a(z)$ is a matrix polynomial in~$z$ of degree~$k$.
  However, the second and third columns in $a(z)$ are
  unaffected by the last factor $G_{k'}(z)$,
  and are consequently only of degree $k-1$.
  To extract the leading coefficient,
  write $G_i(z)=I+z\,(-g_i) (0,0,1)^t (1,0,0)$
  and use $(1,0,0) L_i (0,0,1)^t = l_i^2/2$.
  The constant term is
  \begin{equation*}
    a(0) = L_n L_{n-1} \cdots L_{k'} =
    \begin{pmatrix}
      1 & \sum_{i=k'}^n l_i & \Bigl(\sum_{i=k'}^n l_i \Bigr)^2/2 \\
      0 & 1 & \sum_{i=k'}^n l_i \\
      0 & 0 & 1
    \end{pmatrix}.
  \end{equation*}
  Finally, because $a(z)$ has determinant one,
  its adjoint equals $a(z)^{-1}$,
  which is of degree~$k$ since $G_i(z)^{-1}=G_i(-z)$,
  thus implying that all $2\times 2$ minors of $a(z)$
  are polynomials of degree at most $k$.
\end{proof}

Returning now to equation \eqref{eq:startingpoint},
we divide it by $\phi(1;z)$ and
let $\Phi(y_{k'}-;z)=(p_{k},q_{k},r_{k})^t$
to obtain
\begin{equation}
  \label{eq:evenapprox}
  \begin{pmatrix} 1 \\ W(z) \\ Z(z) \end{pmatrix}
  = \frac{1}{\phi(1;z)} \, a^{(2k)}(z)
  \begin{pmatrix} p_{k}(z) \\ q_{k}(z) \\ r_{k}(z) \end{pmatrix}.
\end{equation}
Suppressing the dependence on $z$ and $k$ in order
to simplify the notation,
we find
\begin{equation}
  \label{eq:Wfraclin}
    \frac{W}{1}
    = \frac{a_{21}\,p+a_{22}\,q+a_{23}\,r}{a_{11}\,p+a_{12}\,q+a_{13}\,r}
    = \frac{a_{21}\,R+a_{22}+a_{23}\,\widehat{R}}{a_{11}\,R+a_{12}+a_{13}\,\widehat{R}},
\end{equation}
where $R=p/q$ and $\widehat{R}=r/q$.
Similarly,
\begin{equation}
  \label{eq:Zfraclin}
  \frac{Z}{1}
  = \frac{a_{31}\,R+a_{32}+a_{33}\,\widehat{R}}{a_{11}\,R+a_{12}+a_{13}\,\widehat{R}}.
\end{equation}
The quantities $R$ and $\widehat{R}$, or more precisely
\begin{equation}
  \label{eq:remainders}
  R_{2k} = \frac{p_{k}}{q_{k}} = \frac{\phi(y_{k'};z)}{\phi_y(y_{k'};z)}
  \qquad\text{and}\qquad
  \widehat{R}_{2k} = \frac{r_{k}}{q_{k}} = \frac{\phi_{yy}(y_{k'}-;z)}
{\phi_y(y_{k'};z)},
\end{equation}
are the analogs of \emph{remainders} referred to earlier.
By formally setting the remainders equal to zero in
\eqref{eq:Wfraclin} and \eqref{eq:Zfraclin}
we get the analogs of \emph{convergents},
$W \approx a_{22}/a_{12}$
and
$Z \approx a_{32}/a_{12}$,
with the following order of approximation:

\begin{theorem}
  \label{thm:approximants}
  As $z\to\infty$,
  \begin{subequations}\label{eq:order}
    \begin{align}
      W(z) &= \frac{a^{(2k)}_{22}(z)}{a^{(2k)}_{12}(z)}+O\left(\frac{1}{z^{k-1}}\right),
      \label{eq:weven}\\
      Z(z) &= \frac{a^{(2k)}_{32}(z)}{a^{(2k)}_{12}(z)}+O\left(\frac{1}{z^{k-1}}\right).
      \label{eq:zeven}
    \end{align}
  \end{subequations}
\end{theorem}

\begin{proof}
  It follows from \eqref{eq:Wfraclin} that
  \begin{equation*}
      W - \frac{a_{22}}{a_{12}} =
      \frac{\ds\frac{a_{12}a_{21}-a_{11}a_{22}}{a_{12}} \,R +
        \frac{a_{12}a_{23}-a_{13}a_{22}}{a_{12}} \,\widehat{R}}{a_{11}\,R+a_{12}+a_{13}\,\widehat{R}}.
  \end{equation*}
  By \eqref{eq:degreePhi}, $p$, $q$, and $r$ are all of degree $k'-1$,
  hence $R=p/q=O(1)$ and $\widehat{R}=r/q=O(1)$.
  From this, and from Lemma~\ref{lem:a-facts}, we see that the right-hand side
  equals
  \begin{equation*}
    \frac{\ds\frac{O(z^{k})}{O(z^{k-1})}O(1) + \frac{O(z^{k})}{O(z^{k-1})}O(1)}
    {O(z^k) O(1) + O(z^{k-1}) + O(z^{k-1}) O(1)}
    = O\left(\frac{1}{z^{k-1}}\right).
  \end{equation*}
  The proof for $Z$ is similar.
\end{proof}

\begin{theorem}
  \label{thm:suppl}
  Let
  \begin{equation}
    W^*(z)=-W(-z),
    \qquad
    Z^*(z)=Z(-z).
  \end{equation}
  Then, with $a=a^{(2k)}(z)$,
  \begin{equation}
    \label{eq:evensuppeq}
    Z^* \, a_{12} + W^* \, a_{22} + a_{32}
    = O\left(\frac{1}{z^k}\right),
    \qquad
    \text{as $z\to\infty$}.
  \end{equation}
\end{theorem}

\begin{proof}
  From \eqref{eq:evenapprox} we have
  \begin{equation*}
    a(z)^{-1}
    \begin{pmatrix} 1\\W(z)\\Z(z) \end{pmatrix}
    =\frac{1}{\phi(1;z)} \Phi(y_{k'}-;z)
    = O\left(\frac{1}{z^k}\right).
  \end{equation*}
  Now recall \eqref{eq:conjugateJ},
  which implies that $( a(z)^{-1} )^t = J a(-z) J$,
  where
  $J=\left(\begin{smallmatrix}
      0&0&1\\0&-1&0\\1&0&0
    \end{smallmatrix}\right)$.
  (Note that $J=J^{-1}=J^t$.)
  Hence
  \begin{equation*}
    J a(-z)^t J \begin{pmatrix} 1\\W(z)\\Z(z) \end{pmatrix}
    = O\left(\frac{1}{z^k}\right),
  \end{equation*}
  which upon changing $z$ to $-z$ and multiplying both sides by $J$
  yields
  \begin{equation*}
    a(z)^t \begin{pmatrix} Z(-z) \\ -W(-z) \\ 1 \end{pmatrix}
    = O\left(\frac{1}{z^k}\right),
  \end{equation*}
  the second row of which is precisely \eqref{eq:evensuppeq}.
\end{proof}

\subsection{An approximation problem}
\label{sec:approx}

Lemma~\ref{lem:a-facts},
Theorem~\ref{thm:approximants},
and Theorem~\ref{thm:suppl}
imply that the entries in the second column of $a=a^{(2k)}(z)$,
\begin{equation*}
  (Q,P,\widehat{P})=(a_{12},a_{22},a_{32}),
\end{equation*}
satisfy the following approximation problem:

\begin{definition}[Pad\'{e}-like approximation problem]
  \label{def:pade}
  Let the functions $W$ and $Z$ be given by
  \begin{equation}
    \label{eq:WZ-laurent}
    \begin{split}
      \frac{W(z)}{z}
      &= \sum_{k=0}^n \frac{b_k}{z-\lambda_k}
      = \sum_{j=0}^{\infty} \frac{\beta_j}{z^{j+1}},
      \\
      \frac{Z(z)}{z}
      &= \sum_{k=0}^n \frac{c_k}{z-\lambda_k}
      = \sum_{j=0}^{\infty} \frac{\gamma_j}{z^{j+1}},
    \end{split}
  \end{equation}
  where
  \begin{equation}
    \label{eq:beta-gamma}
    \beta_j = \sum_{k=0}^n b_k \lambda_k^j,
    \qquad
    \gamma_j = \sum_{k=0}^n c_k \lambda_k^j,
  \end{equation}
  and the $c_k$'s depend on $b_k$'s and $\lambda_k$'s as in
  Corollary~\ref{cor:ck-from-bk}.
  For a given integer $1 \le k \le n$,
  we seek three polynomials
  $(Q,P,\widehat{P})$ of degree $k-1$
  satisfying the following conditions:
  \begin{enumerate}
  \item (Approximation)
    \begin{equation*}
      W=\frac{P}{Q}+O\left(\frac{1}{z^{k-1}}\right),
      \qquad
      Z=\frac{\widehat{P}}{Q}+O\left(\frac{1}{z^{k-1}}\right)
      \qquad
      (z\to\infty).
    \end{equation*}

  \item (Symmetry)
    \begin{equation*}
      Z^* \, Q + W^* \, P + \widehat{P}
      =O\left(\frac{1}{z^k}\right)
      \qquad (z\to\infty),
    \end{equation*}
    where $W^*(z)=-W(-z)$ and $Z^*(z)=Z(-z)$.

  \item (Normalization at $z=0$)
    \begin{equation*}
      P(0)=1,
      \qquad
      \widehat{P}(0)=0.
    \end{equation*}
  \end{enumerate}
\end{definition}

\begin{remark}
  The approximation condition is similar to
  classical Pad\'{e}--Hermite approximation \cite{ns},
  where one approximates two (independently chosen) functions
  to order $O(1/z^{2(k-1)})$
  using a common denominator of degree $k-1$.
  Since in our case we only have $O(1/z^{k-1})$,
  there is clearly not enough information in the approximation
  condition alone to uniquely determine the denominator $Q(z)$.
  However, our $W$ and $Z$ are not independent of each other,
  and the additional symmetry condition
  will eventually lead to a unique $Q(z)$.
\end{remark}

If we introduce \emph{spectral measures}
\begin{equation}
  \label{eq:spectral-measures}
  \mu = \sum_{k=0}^n b_k \, \delta_{\lambda_k}
  \qquad\text{and}\qquad
  \nu = \sum_{k=0}^n c_k \, \delta_{\lambda_k},
\end{equation}
then
\begin{equation}
  \label{eq:beta-gamma2}
  \beta_j = \sum_{k=0}^n b_k \lambda_k^j = \int x^j \, d\mu(x)
  \qquad\text{and}\qquad
  \gamma_j = \sum_{k=0}^n c_k \lambda_k^j = \int x^j \, d\nu(x)
\end{equation}
are the moments of $\mu$ and $\nu$, respectively.
By Corollary~\ref{cor:ck-from-bk}, $\nu$ is determined by $\mu$.
In fact, \eqref{eq:ck} can be expressed as
\begin{equation}
  \label{eq:int-dnu}
  \int f(x)\,d\nu(x)
  = \iint \frac{x\,f(x)}{x+y} \, d\mu(x) \, d\mu(y)=\iint \frac{y\,f(y)}{x+y} \, d\mu(x) \, d\mu(y).
\end{equation}
In particular,
\begin{equation}
  \label{eq:gamma}
  \gamma_j = \iint \frac{x^{j+1}}{x+y} \, d\mu(x) \, d\mu(y)=\iint \frac{y^{j+1}}{x+y} \, d\mu(x) \, d\mu(y).
\end{equation}

\begin{lemma}
  \label{lem:wz-constraint}
  The relation \eqref{eq:int-dnu} between $\mu$ and $\nu$ implies that
  $W$ and $Z$ satisfy the constraint
  \begin{equation*}
    Z(z)+Z^*(z)+W^*(z)W(z)=0.
  \end{equation*}
\end{lemma}

\begin{proof}
  Expressing $W$ and $Z$ in terms of the spectral measures
  we obtain
  \begin{equation*}
    W(z)=\int \frac{z}{z-x} \, d\mu(x),\qquad W^*(z) =
    -\int \frac{z}{z+y} \, d\mu(y),
  \end{equation*}
  and
  \begin{align*}
    Z(z)=\int \frac{z}{z-x} \, d\nu(x)=&\iint \frac{zx}{z-x} \frac{1}{x+y} \, d\mu(x) \, d\mu(y), \\
    Z^*(z)=\int \frac{z}{z+x} \, d\nu(x)=&\iint \frac{zy}{z+y} \frac{1}{x+y} \, d\mu(x) \, d\mu(y),
  \end{align*}
  from which the identity in question follows immediately.
\end{proof}

For the purposes of this approximation problem,
we can consider all functions as being formal Laurent series
with finitely many positive powers.
Let $\Pi_{\pm}$ and $\Pi_{+}^0$ denote the following projection operators
on finite dimensional subspaces
(we regard $k$ as fixed, and omit dependence on $k$ in the notation):
\begin{align*}
&  \Pi_+ \biggl( \sum_{i=-\infty}^{N} a_i z^i \biggr)
  = \sum_{i=1}^{k-1} a_i z^i,
  \qquad
\Pi_+^0\biggl( \sum_{i=-\infty}^{N} a_i z^i \biggr)
  = \sum_{i=0}^{k-1} a_i z^i,\\
  \qquad
&\Pi_- \biggl( \sum_{i=-\infty}^{N} a_i z^i \biggr)
  = \sum_{i=-(k-1)}^{0} a_i z^i.
\end{align*}
Let $M_f$ denote multiplication by $f(z)$,
and define the truncated Hankel and Toeplitz operators
\begin{equation}
  \label{eq:HToperators}
  \begin{split}
    H_f &= \Pi_- \circ M_{f}\circ \Pi_{+}^0 \, : \,
    \linspan\{ z^0,\dots,z^{k-1} \}
    \to
    \linspan\{ z^{-k+1},\dots,z^{0} \},
    \\
    T_f &= \Pi_+ \circ M_f\circ \Pi_{+}^0 \, : \,
    \linspan\{ z^0,\dots,z^{k-1} \}
    \to
    \linspan\{ z^1,\dots,z^{k-1} \}.
  \end{split}
\end{equation}
Moreover, let
\begin{equation*}
  \iota \, : \,
  \linspan\{ z^1,\dots,z^{k-1} \}
  \to
  \linspan\{ z^0,z^1,\dots,z^{k-1} \}
\end{equation*}
be the natural inclusion.

\begin{theorem}\label{thm:linpade}
  The polynomials $Q(z)$, $P(z)$, $\widehat{P}(z)$ solve the approximation
  problem of Definition~\ref{def:pade} if and only if
  \begin{gather}
    \label{eq:P-Phat}
    P=1+ \Pi_+ W Q,
    \qquad
    \widehat{P} = \Pi_+ Z Q, \\
    \label{eq:Qhankel}
    \bigl[ H_{Z^*} + H_{W^*} \circ \iota \circ T_W \bigr] \, Q = -\Pi_- W^*.
  \end{gather}
  Equation \eqref{eq:Qhankel} is equivalent to the linear system
  \begin{equation}
    \label{eq:systemQ}
    \begin{pmatrix}
      \delta_{00} & \delta_{01} & \dots & \delta_{0,k-1} \\
      \delta_{10} & \delta_{11} & \dots & \delta_{1,k-1} \\
      \vdots & \vdots && \vdots \\
      \delta_{k-1,0} & \delta_{k-1,1} & \dots & \delta_{k-1,k-1}
    \end{pmatrix}
    \begin{pmatrix}
      q_0 \\ q_1 \\ \vdots \\ q_{k-1}
    \end{pmatrix}
    =
    \begin{pmatrix}
      \beta_0 \\ \beta_1 \\ \vdots \\ \beta_{k-1}
    \end{pmatrix}
  \end{equation}
  for the unknown coefficients in
  $Q(z)=\sum_{i=0}^{k-1} q_i z^i$,
  where
  \begin{equation}
    \label{eq:delta-ab}
    \delta_{ab}=\iint \frac{x^{a+1}y^b}{x+y} \, d\mu(x) d\mu(y).
  \end{equation}
\end{theorem}

\begin{proof}
  According to the approximation condition,
  the functions $WQ-P$ and $ZQ-\widehat{P}$ contain no
  positive powers of $z$,
  hence their $\Pi_+$ projections are zero.
  This, together with the normalization conditions,
  is equivalent to \eqref{eq:P-Phat}
  (which shows that $P$ and $\widehat{P}$ are uniquely determined by $Q$).
  Inserting \eqref{eq:P-Phat} into the symmetry condition yields
  \begin{equation}
    \label{eq:symmcondQ}
    Z^* \, Q
    + W^* \, \bigl(1 + \Pi_+ W Q \bigr)
    + \Pi_+ Z Q
    = O\left(\frac{1}{z^k}\right).
  \end{equation}
  The coefficients of the positive powers
  $z^1,\dots,z^{k-1}$ appearing in \eqref{eq:symmcondQ} vanish
  identically regardless of $Q$;
  indeed,
  using $(\Pi_+-I)Q=-q_0$, $\Pi_+ W^*=0$,
  and Lemma~\ref{lem:wz-constraint},
  we obtain
  \begin{multline*}
    \Pi_+ \Bigl(
    Z^* \, Q
    + W^* \, \bigl(1 + \Pi_+ W Q \bigr)
    + \Pi_+ Z Q
    \Bigr)
    =
    \Pi_+\Bigl(Z+Z^*+W^*\Pi_+W \Bigr) Q
    \\
    = \Pi_+\Bigl(Z+Z^*+W^*(\Pi_+-I+I)W\Bigr)Q
    =\Pi_+\Bigl(Z+Z^*+W^*W\Bigr)Q
    =0.
  \end{multline*}
  On the other hand,
  the vanishing of the coefficients of $z^{-(k-1)},\dots,z^0$
  in \eqref{eq:symmcondQ} is equivalent to the following
  condition on $Q$:
  \begin{equation}
    \label{eq:symmcondQproj-}
    \Pi_- \Bigl( {Z^*} \, Q
    + {W^*} \, \bigl(1 + \Pi_+ W Q \bigr)
    + \Pi_+ Z Q \Bigr)
    = 0,
  \end{equation}
  which is just \eqref{eq:Qhankel}.

  To show equivalence of
  \eqref{eq:Qhankel} and \eqref{eq:systemQ} we write out
  \eqref{eq:Qhankel} in terms of matrices
  with respect to the standard bases ordered as in
  \eqref{eq:HToperators}.
  Then $Q$ is represented by the column vector
  $(q_0,\dots,q_{n-1})^t$,
  while
  \begin{equation*}
    \begin{split}
      H_{Z^*} &=\Bigl[ (-1)^{a+b} \gamma_{a+b} \Bigr]_{\substack{0 \le a \le k-1\\0 \le b \le k-1}} =
      \begin{pmatrix}
        \gamma_0 & -\gamma_1 & \gamma_2 & \cdots \\
        -\gamma_1 & \gamma_2 && \\
        \gamma_2 &&& \\
        \vdots &&&
      \end{pmatrix},
      \\
      H_{W^*} \circ \iota &=\Bigl[ (-1)^{a+m+1}\beta_{a+m} \Bigr]_{\substack{0 \le a \le k-1\\1 \le m \le k-1}} =
      \begin{pmatrix}
        \beta_1 & -\beta_2 & \beta_3 & \cdots \\
        -\beta_2 & \beta_3 && \\
        \beta_3 &&& \\
        \vdots &&&
      \end{pmatrix},
      \\
      T_W &=\Bigl[ \beta_{b-m} \Bigr]_{\substack{1 \le m \le k-1\\0 \le b \le k-1}} =
      \begin{pmatrix}
        0 & \beta_0 & \beta_1 & \beta_2 & \cdots \\
        0 & 0 & \beta_0 & \beta_1 & \ddots \\
        0 & 0 & 0 & \beta_0 & \ddots \\
        \vdots &&&& \ddots
      \end{pmatrix},
      \\
      -\Pi_- W^* &=
      (\beta_0,-\beta_1,\beta_2,\dots,(-1)^{k-1} \beta_{k-1})^t.
    \end{split}
  \end{equation*}
  If we multiply out the matrices and change the sign of every second row,
  the system takes the form
  \begin{equation}\label{eq:compsystemQ}
    \sum_{b=0}^{k-1}\Bigl[ (-1)^b \gamma_{a+b}-
    \sum_{m=1}^b (-1)^m\beta_{a+m}\beta_{b-m} \Bigr] q_b
    =\beta_a , \quad 0\le a\le k-1.
  \end{equation}
  (When $b=0$, the inner sum is empty.)
  Upon using the spectral representations \eqref{eq:beta-gamma2}
  and \eqref{eq:gamma} of $\beta$'s and $\gamma$'s
  we arrive at the claimed integral representation
  \eqref{eq:delta-ab}
  of the $(a,b)$ entry $\delta_{ab}$ of the matrix of coefficients in
  \eqref{eq:compsystemQ}:
  \begin{multline*}
    \delta_{ab} =
    \iint \Bigl[ (-1)^b \frac{x^{a+b+1}}{x+y}-\sum_{m=1}^b (-1)^m x^{a+m}y^{b-m}\Bigr]
    \, d\mu(x)\, d\mu(y)
    \\
    = \iint \frac{x^{a+1}y^b}{x+y} \, d\mu(x) \, d\mu(y).
  \end{multline*}
\end{proof}

To complete the picture,
it remains to show that the system \eqref{eq:systemQ}
has nonzero determinant,
so that the solution to the approximation problem exists and is unique.
This will be dealt with in the following sections.

\begin{remark}
  The choice of convergents and remainders is not unique.
  For example, one could take the entries in the
  \emph{first} column of $a^{(2k)}(z)$
  as convergents, and instead of \eqref{eq:Wfraclin} write
  \begin{equation*}
    W = \frac{\ds a_{21} + a_{22} \, \frac{q}{p} + a_{13} \, \frac{r}{p}}
    {\ds a_{11} + a_{12} \, \frac{q}{p} + a_{13} \, \frac{r}{p}}
    \approx
    \frac{a_{21}}{a_{11}},
  \end{equation*}
  where now $q/p$ and $r/p$ play the role of remainders,
  and the order of approximation can be shown to be $O(1/z^k)$;
  similarly with $Z$ and equation \eqref{eq:Zfraclin}.
  In this way one obtains an approximation problem which is similar to,
  but slightly different from, the one in Definition~\ref{def:pade}.
  When looking at the first column, the odd matrices $a^{(2k-1)}$
  are different from the even ones,
  and this case provides yet another approximation problem of similar type.
  However, we are convinced that the problem we have chosen to focus
  on here is the simplest route to the solution of the inverse problem.
\end{remark}

\subsection{Determinants}

From the expression \eqref{eq:delta-ab} for $\delta_{ab}$ as a double
integral, it is clear that the determinants that appear when one tries
to solve the linear system \eqref{eq:systemQ} with Cramer's rule
are somewhat reminiscent
of the classical $k \times k$ Hankel determinant of moments,
\begin{equation}
  \label{eq:Hankel-Heine}
  \begin{vmatrix}
    \beta_0 & \beta_1 & \dots & \beta_{k-1} \\
    \beta_1 & \beta_2 & \dots & \beta_{k} \\
    \vdots &&& \vdots \\
    \beta_{k-1} & \beta_{k} & \dots & \beta_{2k-2} \\
  \end{vmatrix}
  =
  \frac{1}{k!} \int_{\R^k} \Delta(x)^2
  d\mu^k(x),
\end{equation}
where $\Delta(x)=\Delta(x_1,\dots,x_k)=\prod_{i<j}(x_i-x_j)$.
(For a proof of Heine's formula \eqref{eq:Hankel-Heine},
see for instance \cite[page 27] {Szego} or \cite[Proposition~3.8]{deift}.)

The following lemma contains similar integral formulas for the
present determinants.
These formulas are valid for any measure,
although our main interest here is of course in
the discrete case when
$\mu=\sum_{i=0}^n b_i \, \delta_{\lambda_i}$.

\begin{lemma}
  \label{lem:super-Heine}
  Suppose $\mu$ is a measure on $\R$ such that the integrals
  \begin{equation}
    \label{eq:beta-delta}
    \beta_a = \int x^a \, d\mu(x)
    \qquad\text{and}\qquad
    \delta_{ab} =
    \iint \frac{x^{a+1} y^{b}}{x+y} \,d\mu(x) \, d\mu(y)
  \end{equation}
  are finite. Let
  \begin{equation}
    \label{eq:integrals-uv}
    \begin{split}
      u_k &=
      \frac{1}{k!}
      \int_{\R^k} \frac{\Delta(x)^2}{\Gamma(x)} d\mu^k(x),
      \\
      v_k &=
      \frac{1}{k!}
      \int_{\R^k} \frac{\Delta(x)^2}{\Gamma(x)} \, x_1 x_2 \cdots x_k \, d\mu^k(x),
    \end{split}
  \end{equation}
  where
  \begin{equation}
    \begin{split}
      \Delta(x)&=\Delta(x_1,\dots,x_k)=\prod_{i<j}(x_i-x_j),\\
      \Gamma(x)&=\Gamma(x_1,\dots,x_k)=\prod_{i<j}(x_i+x_j).
    \end{split}
  \end{equation}
  (When $k=0$ or~$1$, we let $\Delta(x)=\Gamma(x)=1$.
  Also, $u_0=v_0=1$.)
  Then for $k \ge 1$ the following $k\times k$ determinant formulas hold:
  \begin{equation}
    \label{eq:super-Heine1}
    D_k :=
    \begin{vmatrix}
      \delta_{00} & \delta_{01} & \delta_{02} & \dots & \delta_{0,k-1} \\
      \delta_{10} & \delta_{11} & \delta_{12} & \dots & \delta_{1,k-1} \\
      \delta_{20} & \delta_{21} & \delta_{22} & \dots & \delta_{2,k-1} \\
      \vdots & \vdots & \vdots && \vdots \\
      \delta_{k-1,0} & \delta_{k-1,1} & \delta_{k-1,2} & \dots & \delta_{k-1,k-1}
    \end{vmatrix}
    = \frac{(u_k)^2}{2^k},
  \end{equation}
  \begin{equation}
    \label{eq:super-Heine2}
    D'_k :=
    \begin{vmatrix}
      \beta_0 & \delta_{00} & \delta_{01} & \dots & \delta_{0,k-2} \\
      \beta_1 & \delta_{10} & \delta_{11} & \dots & \delta_{1,k-2} \\
      \beta_2 & \delta_{20} & \delta_{21} & \dots & \delta_{2,k-2} \\
      \vdots & \vdots & \vdots && \vdots \\
      \beta_{k-1} & \delta_{k-1,0} & \delta_{k-1,1} & \dots & \delta_{k-1,k-2}
    \end{vmatrix}
    = \frac{u_k \, u_{k-1}}{2^{k-1}},
  \end{equation}
  and
  \begin{equation}
    \label{eq:super-Heine3}
    D''_k :=
    \begin{vmatrix}
      \beta_0 & \delta_{01} & \delta_{02} & \dots & \delta_{0,k-1} \\
      \beta_1 & \delta_{11} & \delta_{12} & \dots & \delta_{1,k-1} \\
      \beta_2 & \delta_{21} & \delta_{22} & \dots & \delta_{2,k-1} \\
      \vdots & \vdots & \vdots && \vdots \\
      \beta_{k-1} & \delta_{k-1,1} & \delta_{k-1,2} & \dots & \delta_{k-1,k-1}
    \end{vmatrix}
    = \frac{u_k \, v_{k-1}}{2^{k-1}}.
  \end{equation}
\end{lemma}

\begin{proof}
  We show only \eqref{eq:super-Heine2};
  the formulas \eqref{eq:super-Heine1} and \eqref{eq:super-Heine3}
  are proved similarly.
  Notationally it is slightly easier to work with
  the $(k+1)\times(k+1)$ determinant $D'_{k+1}$,
  so what we will actually prove is that
  \begin{equation}
    \label{eq:super-Heine2plus}
    D'_{k+1} = \frac{u_k \, u_{k+1}}{2^{k}}.
  \end{equation}

  Note first that our definition of $\Delta(x)$ is such that
  the Vandermonde determinant equals
  \begin{equation*}
    \nonumber
    \det(x_i^{j-1})_{i,j=1,\dots,k}
    = \prod_{i>j}(x_i-x_j)
    = \Delta(x_k,\dots,x_1)
    = (-1)^{\binom{k}{2}} \Delta(x_1,\dots,x_k).
  \end{equation*}
  In the final result the sign convention does not matter since only
  $\Delta(x)^2$ appears there,
  but we need to keep track of the correct signs during the proof.
  As regards $\Gamma(x)$, the order of the variables is of course immaterial.

  Like in the proof of Heine's formula, we use different dummy variables
  in each column, in order to be able to pull integrals outside of the
  determinant by multilinearity.
  We will need $2k+1$ variables $(x_0,x_1,\dots,x_{2k})$, as follows.
  In the integrals $\beta_a$ in the first column we
  call the integration variable $x_0$,
  in the double integrals $\delta_{a0}$ in the second column we use
  $x_1$ and $x_2$,
  in the double integrals $\delta_{a1}$ in the third column we use
  $x_3$ and $x_4$,
  etc.,
  with even-numbered variables replacing $x$
  and odd-numbered variables replacing $y$.
  Then we can pull out all integral signs,
  and also the factor $x_{2b-1}^{b-1} x_{2b} / (x_{2b-1}+x_{2b})$
  from column $b+1$, for $b=1,\dots,k$.
  What remains is a Vandermonde determinant in the even-numbered variables,
  so
  \begin{equation*}
    D'_{k+1}=
    \int_{\R^{2k+1}}
    \frac{(x_1^0 x_3^1 x_5^2 \cdots x_{2k-1}^{k-1})(x_2 x_4 \cdots x_{2k})
      \, \Delta(x_{2k},\dots,x_4,x_2,x_0)}
    {(x_1+x_2)(x_3+x_4) \cdots (x_{2k-1}+x_{2k})}
    \, d\mu^{2k+1}(x).
  \end{equation*}
  We now introduce an auxiliary variable $x_{-1}=0$ in order to write
  the last expression in a more symmetric form:
  \begin{equation*}
    D'_{k+1}=
    \int_{\R^{2k+1}}
    \frac{(x_1^0 x_3^1 x_5^2 \cdots x_{2k-1}^{k-1})(x_0x_2 x_4 \cdots x_{2k})
      \, \Delta(x_{2k},\dots,x_4,x_2,x_0)}
    {(x_{-1}+x_0)(x_1+x_2)(x_3+x_4) \cdots (x_{2k-1}+x_{2k})}
    \, d\mu^{2k+1}(x).
  \end{equation*}

  The value of the original determinant is independent of
  the order in which we number the dummy variables,
  so the above expression is invariant under permutations of the
  variables.
  In particular, symmetrizing over the even-numbered variables
  doesn't change the value.
  That is, if $S_{\text{even}}$ denotes the set of permutations of
  $\{0,2,4,\dots,2k\}$, then
  \begin{multline*}
    D'_{k+1}=
    \frac{1}{(k+1)!}
    \int_{\R^{2k+1}}
    (x_1^0 x_3^1 \cdots x_{2k-1}^{k-1})(x_0x_2 x_4 \cdots x_{2k})
    \,
    \Delta(x_{2k},\dots,x_4,x_2,x_0)
    \\
    \times
    \left(
      \sum_{\pi\in S_{\text{even}}}
      \frac{(\sgn\pi)}
      {(x_{-1}+x_{\pi(0)})(x_1+x_{\pi(2)})(x_3+x_{\pi(4)})
        \cdots (x_{2k-1}+x_{\pi(2k)})}
    \right)
    \, d\mu^{2k+1}(x).
  \end{multline*}
  The sum in parentheses is a Cauchy determinant,
  which by the well-known formula equals
  \begin{multline*}
    \frac{\Delta (x_{-1},x_1,x_3,\dots,x_{2k-1})\Delta(x_0,x_2,x_4,\cdots,x_{2k})}
    {\prod_{i=0}^k \prod_{j=0}^k (x_{2i-1}+x_{2j})}
    \\
    =\frac{(-x_1)(-x_3)\cdots (-x_{2k-1})\Delta (x_1,x_3,\dots,x_{2k-1})\Delta(x_0,x_2,x_4,\cdots,x_{2k})}
    {(x_0x_2 x_4 \cdots x_{2k}) \prod_{i=1}^k \prod_{j=0}^k (x_{2i-1}+x_{2j})},
  \end{multline*}
  where the second line is a result of splitting off terms involving $x_{-1}$.
  It follows that
  \begin{multline*}
    D'_{k+1}=
    \frac{1}{(k+1)!}
    \int_{\R^{2k+1}}(-1)^{\binom{k+1}{2}+k}(x_1x_3 \cdots x_{2k-1})
    (x_1^0 x_3^1 \cdots x_{2k-1}^{k-1})
    \\
    \times \frac{\Delta(x_1,x_3,\dots, x_{2k-1})\Delta(x_0,x_2,\dots,x_{2k})^2}
    {\prod_{i=1}^k \prod_{j=0}^k (x_{2i-1}+x_{2j})}
    \, d\mu^{2k+1}(x).
  \end{multline*}
  Symmetrizing this over the odd-numbered variables we get yet
  another Vandermonde determinant
  $(-1)^{\binom{k}{2}} \Delta(x_1,x_3,\dots,x_{2k-1})$,
  hence
  \begin{multline*}
    D'_{k+1}=
    \frac{1}{k! (k+1)!}
    \int_{\R^{2k+1}}
    (x_1x_3 \cdots x_{2k-1})
    \\
    \times \frac{\Delta(x_1,x_3,\dots,x_{2k-1})^2 \,
      \Delta(x_0,x_2,\dots,x_{2k})^2}
    {\prod_{i=1}^k \prod_{j=0}^k (x_{2i-1}+x_{2j})}
    \, d\mu^{2k+1}(x),
  \end{multline*}
  and then symmetrizing over all variables gives
  \begin{multline*}
    D'_{k+1}=
    \frac{1}{k! (k+1)! (2k+1)!}
    \int_{\R^{2k+1}}
    \biggl(
    \sum_{\pi\in S_{2k+1}}(x_{\pi(1)}x_{\pi(3)} \cdots x_{\pi(2k-1)})\\
    \times \frac{ \Delta(x_{\pi(1)},x_{\pi(3)},\dots,x_{\pi(2k-1)})^2 \,
      \Delta(x_{\pi(0)},x_{\pi(2)},\dots,x_{\pi(2k)})^2}
    {\prod_{i=1}^k \prod_{j=0}^k (x_{\pi(2i-1)}+x_{\pi(2j)})}
    \biggr)
    \, d\mu^{2k+1}(x).
  \end{multline*}
  If $\sigma\in S_{2k+1}$ leaves the sets $\{0,2,\dots,2k \}$ and
  $\{ 1,3,\dots,2k-1 \}$ invariant,
  then the terms corresponding to $\pi$ and to $\pi\circ\sigma$
  are equal,
  so each term actually appears $k!(k+1)!$ times in the sum above.
  Using the notation
  $\binom{[0,n]}{k}$
  for the set of $k$-element subsets
  $\{ i_1<\dots<i_k \}$ of $\{ 0, 1,\dots,n \}$,
  as well as other notation introduced earlier in
  Definition~\ref{def:notation-galore},
  we can write $D'_{k+1}$ in a less redundant way by picking out a
  ``sorted representative'' from each such group of equal terms:
  \begin{equation*}
    \begin{split}
      D'_{k+1} &=
      \frac{1}{(2k+1)!}
      \int_{\R^{2k+1}}
      \biggl(
      \sum_{I,J}
      x_I \frac{\Delta_I^2 \,
        \Delta_J^2}
      {\Gamma_{I,J}}
      \biggr)
      \, d\mu^{2k+1}(x)
      \\
      &=
      \frac{1}{(2k+1)!}
      \int_{\R^{2k+1}}
      \frac{1}{\Gamma_{I\cup J}}  \biggl(
      \sum_{I,J}
      x_I \Delta_I^2 \,
      \Delta_J^2
      \Gamma_{I}\Gamma_{J}
      \biggr)
      \, d\mu^{2k+1}(x),
    \end{split}
  \end{equation*}
  where the sum runs over all $\binom{2k+1}{k}$ ways of partitioning
  $\{ 0,1,\dots,2k \} = I \cup J$
  into disjoint sets
  $I\in\binom{[0,2k]}{k}$
  and
  $J\in\binom{[0,2k]}{k+1}$.
  Now we claim that the symmetric polynomial given by the sum
  in parentheses satisfies the identity
  \begin{equation*}
    \sum_{I,J}
    x_I \Delta_I^2 \,
    \Delta_J^2
    \Gamma_{I}\Gamma_{J}=\frac{1}{2^k} \sum_{I,J}
    \Delta_I^2 \,
    \Delta_J^2
    \Gamma_{I,J}
  \end{equation*}
  with $I$ and $J$ running over the same sets as before.
  The general strategy for proving identities of this type is
  outlined in Appendix~\ref{sec:app-identities}, which also contains other
  identities needed in the proof of the remaining two
  statements \eqref{eq:super-Heine1} and \eqref{eq:super-Heine3}
  of the present theorem.
  Granted the claim, we obtain
  \begin{equation*}
    \begin{split}
      D'_{k+1} &=
      \frac{2^{-k}}{(2k+1)!}
      \int_{\R^{2k+1}}
      \biggl(
      \sum_{I,J}
      \frac{\Delta_I^2 \, \Delta_J^2}{\Gamma_I \, \Gamma_J}
      \biggr)
      \, d\mu^{2k+1}(x)
      \\
      &=
      \frac{2^{-k} k! (k+1)!}{(2k+1)!}
      \sum_{I,J}
      \biggl(
      \frac{1}{k!}
      \int_{\R^k}
      \frac{\Delta_I^2}{\Gamma_I}
      \, d\mu^{k}(x_{i_1},\dots,x_{i_k})
      \\
      &\qquad\qquad\qquad \times
      \frac{1}{(k+1)!}
      \int_{\R^{k+1}}
      \frac{\Delta_J^2}{\Gamma_J}
      \, d\mu^{k+1}(x_{j_1},\dots,x_{j_{k+1}})
      \biggr)
      \\
      &=
      \frac{2^{-k}}{\binom{2k+1}{k}}
      \sum_{I,J} u_k \, u_{k+1}
      = \frac{u_k \, u_{k+1}}{2^k},
    \end{split}
  \end{equation*}
  which proves \eqref{eq:super-Heine2plus}.
\end{proof}

Here we need yet some more notation.
Let $\wt{V}_{-1}=0$, $\wt{U}_0=\wt{V}_0=1$, and for $k\ge 1$:
\begin{equation}
  \label{eq:UVtilde}
  \wt{U}_k = \sum_{I\in\binom{[0,n]}{k}}
  \frac{(\Delta_{I})^2}{\Gamma_I} b_I,
  \qquad
  \wt{V}_k = \sum_{I\in\binom{[0,n]}{k}}
  \frac{(\Delta_{I})^2}{\Gamma_I} \lambda_I b_I,
\end{equation}
In other words,
$\wt{U}_k$ and $\wt{V}_k$ are symmetric functions of the same
form as $U_k$ and $V_k$
of Definition~\ref{def:notation-galore}, but involving the
additional variables $\lambda_0$ and $b_0$.
For future use we also define, for $k\ge 0$,
\begin{equation}
  \label{eq:Wtilde}
  \wt{W}_k =
  \begin{vmatrix}
    \wt{U}_k & \wt{V}_{k-1} \\
    \wt{U}_{k+1} & \wt{V}_k
  \end{vmatrix}.
\end{equation}

\begin{lemma}
  When $\mu=\sum_{i=0}^n b_i \, \delta_{\lambda_i}$,
  the integrals $u_k$ and $v_k$ in \eqref{eq:integrals-uv}
  reduce to the sums $\wt{U}_k$ and $\wt{V}_k$, respectively.
  Hence, in this case,
  \begin{equation}
    \label{eq:determinants-discrete}
    D_k = \frac{\bigl( \wt{U}_k \bigr)^2}{2^k},
    \qquad
    D'_k = \frac{\wt{U}_k \, \wt{U}_{k-1}}{2^{k-1}},
    \qquad
    D''_k = \frac{\wt{U}_k \, \wt{V}_{k-1}}{2^{k-1}}.
  \end{equation}
\end{lemma}

\begin{proof}
  This follows from the definitions and Lemma~\ref{lem:super-Heine}.
\end{proof}

Recall that in the context of the discrete cubic string
we have $\lambda_0=0$ and $b_0=1$ by definition.

\begin{lemma}
  When $\lambda_0=0$ and $b_0=1$,
  \begin{equation}
    \label{eq:UVWtilde-when-01}
    \wt{U}_k = U_k + V_{k-1},
    \qquad
    \wt{V}_k = V_k,
    \qquad
    \wt{W}_k = W_k.
  \end{equation}
\end{lemma}

\begin{proof}
  In the sum defining $\wt{U}_k$,
  the terms for which $0 \notin I$ add up to $U_k$,
  while if $I=\{0\} \cup J$ with $J\in\binom{[1,n]}{k-1}$,
  then $(\Delta_I)^2/\Gamma_I=\lambda_J (\Delta_J)^2/\Gamma_J$ and $b_I=b_J$,
  so those terms add up to $V_{k-1}$.
  For $\wt{V}_k$, the terms with $0\in I$ vanish,
  leaving only terms which add up to $V_k$.
  Finally,
  \begin{equation*}
    \wt{W}_k
    =
    \begin{vmatrix}
      U_k + V_{k-1} & V_{k-1} \\
      U_{k+1} + V_k & V_k
    \end{vmatrix}
    =
    \begin{vmatrix}
      U_k & V_{k-1} \\
      U_{k+1} & V_k
    \end{vmatrix}
  = W_k.
  \end{equation*}
\end{proof}

\subsection{Solution of the inverse problem}
\label{sec:solution-invprob}

Now we merely have to put the pieces together.

\begin{theorem}
  The approximation problem of Definition~\ref{def:pade}
  has a unique solution:
  \begin{enumerate}
  \item The coefficients of the polynomial
    $Q(z)=\sum_{i=0}^{k-1} q_i z^i$
    are uniquely determined by the nonsingular linear system
    \eqref{eq:systemQ}.
    In particular,
    the highest and lowest coefficients of $Q(z)$ are given by the
following ratios of determinants:
    \begin{equation}
      \label{eq:Qcoeff}
      (-1)^{k-1} q_{k-1} = \frac{D'_k}{D_k},     \qquad
      q_0 = \frac{D''_k}{D_k}.
    \end{equation}
  \item The polynomials $P(z)$ and $\widehat{P}(z)$ are
    uniquely determined by $Q(z)$ as
    \begin{equation}
      \label{eq:P-Phat2}
      P = 1 + T_W \, Q,
      \qquad
      \widehat{P} = T_Z \, Q.
    \end{equation}
  \end{enumerate}
\end{theorem}

\begin{proof}
  The determinant of \eqref{eq:systemQ} is
  $D_k = \bigl( \wt{U}_k \bigr)^2/2^k
  = \bigl( U_k + V_{k-1} \bigr)^2/2^k$,
  by the lemmas in the previous section.
  This is clearly positive since $U_k >0$ and $V_{k-1}>0$,
  so the system is nonsingular.
  Solving \eqref{eq:systemQ} for $q_0$ and $q_{k-1}$ using Cramer's rule,
  one obtains the $k\times k$ determinants in
  Lemma~\ref{lem:super-Heine},
  the factor $(-1)^{k-1}$ appearing when moving the $\beta_i$'s
  from the last column to the first
  in the case of $q_{k-1}$.
  Equation \eqref{eq:P-Phat2} is just a reformulation of \eqref{eq:P-Phat}.
\end{proof}

Expressing the result in terms of $\wt{U}$'s and $\wt{V}$'s
using \eqref{eq:determinants-discrete}
yields
\begin{corollary}
  \begin{equation}
    \label{eq:Quvcoeff}
    (-1)^{k-1} q_{k-1} = \frac{2\,\wt{U}_{k-1}}{\wt{U}_k},
    \qquad
    q_0 = \frac{2\,\wt{V}_{k-1}}{\wt{U}_k}.
  \end{equation}
\end{corollary}

\begin{remark}
  Note that there is a factor of $\wt{U}_k$ which cancels in the quotients
  \eqref{eq:Qcoeff}.
  As should be clear from the proof of Lemma~\ref{lem:super-Heine},
  this is a considerable complication compared to the classical case
  (Stieltjes continued fractions, Pad{\'e} approximation,
  ordinary discrete string, Camassa--Holm peakons).
\end{remark}

Returning to the discrete cubic string,
recall that $(Q,P,\widehat{P})=(a_{12},a_{22},a_{32})$
is a solution of our approximation problem.
Comparing equation \eqref{eq:a12} for $a_{12}(z)$
to equation \eqref{eq:Quvcoeff} for $Q(z)$, we see that
\begin{equation}
  \label{eq:magic-wand}
  \begin{split}
    \sum_{i=k'}^n l_i
    &=
    \frac{2\,\wt{V}_{k-1}}{\wt{U}_k}
    =
    \frac{2\,V_{k-1}}{U_k+V_{k-1}},
    \\
    \Omega_k :=
    l_{k'} \left( \prod_{i=k'+1}^n \frac{g_i l_i^2}{2} \right)
    &=
    \frac{2\,\wt{U}_{k-1}}{\wt{U}_k}
    =
    \frac{2\,(U_{k-1}+V_{k-2})}{U_k+V_{k-1}}.
  \end{split}
\end{equation}
Given the quantities in \eqref{eq:magic-wand} one can easily
compute the original string variables from
\begin{equation*}
  y_{k'} = 1-\sum_{i=k'}^n l_i,
  \qquad
  l_{k'} =
  y_{(k-1)'}-y_{k'},
  \qquad
  g_{k'} = \frac{2\,\Omega_{k+1}}{l_{(k+1)'} l_{k'} \Omega_{k}},
\end{equation*}
with the following result:

\begin{theorem}
  \label{thm:solution-inverse-string}
  The solution to the inverse spectral problem for the
  discrete cubic string in terms of symmetric functions
  of $\lambda$'s and $b$'s is
  \begin{equation}
    \label{eq:solution-inverse-string}
    \begin{split}
      y_{k'} &= 1 - \frac{2 \, \wt{V}_{k-1}}{\wt{U}_k}
      = \frac{U_k - V_{k-1}}{U_k + V_{k-1}}
      \qquad (k=1,\dots,n),
      \\
      l_{k'} &= \frac{2 \, \wt{W}_{k-1}}{\wt{U}_{k-1} \wt{U}_k}
      = \frac{2 \, W_{k-1}}{(U_{k-1}+V_{k-2})(U_k+V_{k-1})}
      \qquad (k=1,\dots,n+1),
      \\
      g_{k'} &= \frac{{(\wt{U}_k)^4}}{2 \, \wt{W}_{k-1} \wt{W}_k}
      = \frac{(U_k+V_{k-1})^4}{2 \, W_{k-1} W_k}
      \qquad (k=1,\dots,n),
    \end{split}
  \end{equation}
  with $U_k$, $V_k$, and $W_k$ as in Definition~\ref{def:notation-galore}.
\end{theorem}

\begin{remark}
  The formula for $y_{k'}$ is valid also for $k=0$ and $k=n+1$;
  since $V_0=U_{n+1}=0$,
  it yields $y_{n+1}=+1$ and $y_0=-1$, respectively, as it should.
\end{remark}

Now we can finally prove the remaining part of
Theorem~\ref{thm:inverse-problem-peakon},
namely the explicit formulas for the inverse spectral problem
in terms of peakon variables $x_k$ and $m_k$.

\begin{lemma}
  \begin{equation}
    \label{eq:n-peakon-solution-again}
    x_{k'} = \log \frac{U_k}{V_{k-1}},
    \qquad
    m_{k'} =
    \frac{(U_k)^2 \, (V_{k-1})^2}{W_k W_{k-1}}
    \qquad
    (k=1,\dots,n).
  \end{equation}
\end{lemma}

\begin{proof}
  By \eqref{eq:measure-g} the string variables can be expressed
  in terms of peakon variables as
  \begin{equation}
    \label{eq:change-of-vars-again}
    x_{k'}=\log \frac{1+y_{k'}}{1-y_{k'}},
    \qquad
    m_{k'} = \frac{1}{8} g_{k'} (1-y_{k'}^2)^2,
  \end{equation}
  which immediately yields
  \eqref{eq:n-peakon-solution-again}
  upon inserting \eqref{eq:solution-inverse-string}.
\end{proof}

So far the solution of the inverse problem has been obtained
under the assumption
that the spectral measure is coming from a cubic string.
Theorem~\ref{thm:inverse-problem-peakon} implies however that there is a
bijection between discrete cubic strings and spectral measures $\mu$.

\begin{theorem}
  \label{thm:inverse-problem-solved}
  For every discrete cubic string specified by a positive measure
  $g(y)=\sum_{i=1}^n g_i \, \delta_{y_i} (y)$
  whose support satisfies $-1<y_1<y_2<\cdots<y_n <1$ there exists
  a unique positive measure
  \begin{equation}\label{def:mumeasure}
    \mu=\sum_{i=0}^n b_i \, \delta_{\lambda_i},
  \end{equation}
  $b_0=1$, $b_k>0$,
  $0=\lambda_0<\lambda_1<\cdots<\lambda_n$,
  such that the Weyl function $W(z)$ of the cubic string satisfies
  \begin{equation*}
    \frac {W(z)}{z}
    =\int \frac{d\mu(\lambda)}{z-\lambda}.
  \end{equation*}
  Conversely, given a measure $\mu$ as in \eqref{def:mumeasure} there
  exists a unique cubic string defined by
  Theorem~\ref{thm:solution-inverse-string}.
\end{theorem}

\section{Acknowledgments}

The research leading up to this work was initiated while the first
author was visiting the Department of Mathematics and Statistics of
the University of Saskatchewan, Canada, in the academic year
2002/2003. The research of the first author has been supported in part
by the Department of Mathematics and Statistics, University of
Saskatchewan, and by the National Science and Engineering Research
Council of Canada (NSERC).  The research of the second author is
supported by NSERC.  Both authors would like to thank NSERC and the
Department of Mathematics and Statistics of the University of
Saskatchewan for making this collaboration possible.

\appendix
\section{Appendix: The discrete string}
\label{sec:app-string}

Here we give a brief account of the solution of the inverse spectral problem
for a discrete string with Dirichlet boundary condition,
for comparison to the discrete \emph{cubic} string treated in this paper.
For more details, we refer the reader to \cite{bss-moment},
and to Stieltjes' famous
\emph{Recherches sur les fraction continues},
especially the introduction and Chapter~3
\cite[Vol.~II]{stieltjes}.
It was M.~G.~Krein who pointed out the
mechanical interpretation of Stieltjes' work
and generalized it in his spectral theory of the
general (not necessarily discrete) inhomogenous string
\cite{kackrein}; see also Supplement II in \cite{gantmacher-krein}, or
the comprehensive treatment in \cite{McKean}.

Small vibrations $u(y,t)$ of a string with mass density $g(y)$
are described by the wave equation $u_{yy}=g(y) u_{tt}$.
If the ends of the string are attached at $y=\pm 1$,
separation of variables $u(y,t)=\phi(y)\tau(t)$
results in the following equation for $\phi(y)$:
\begin{equation}
  \label{eq:dstring-spectral}
  \begin{gathered}
    \phi_{yy}(y) = z \, g(y) \, \phi(y)
    \quad\text{for $y \in (-1,1)$},
    \\
    \phi(-1) = 0,
    \qquad
    \phi(1) = 0.
  \end{gathered}
\end{equation}
This is a selfadjoint spectral problem with simple, negative (if $g>0$)
eigenvalues $z=\lambda_k=-\omega_k^2$,
where $\omega_k$ is the frequency of the $k$th harmonic.
The \emph{discrete} string consisting of point masses~$g_i$
at positions $-1<y_1<\cdots<y_n<+1$ corresponds to
$g(y)=\sum_1^n g_i \delta(y-y_i)$
being a discrete measure.
In this case,
there are exactly $n$ eigenvalues,
and the eigenfunctions $\phi(y)$ are piecewise linear
(as opposed to the well known sinusoidal shape in the
homogeneous case $g(y)\equiv 1$).
The piecewise constant slope $\phi_y(y)$ jumps by
$z \, g_i \, \phi(y_i)$ at each point $y_i$.

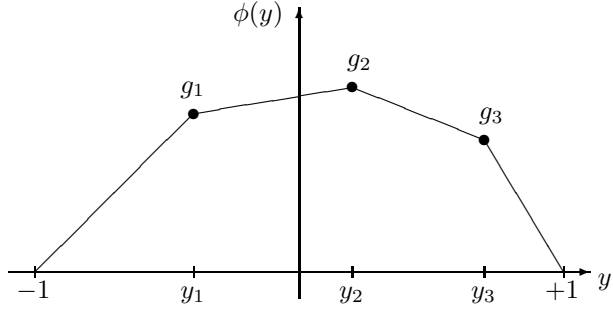
\begin{figure}[h!]
  \begin{center}
    \begin{picture}(250,110)
    \put(0,10){\vector(1,0){220}} \put(223,5){$y$}
    \put(110,0){\vector(0,1){110}} \put(85,105){$\phi(y)$}
    \put(10,8){\line(0,1){4}} \put(3,0){$-1$}
    \put(70,8){\line(0,1){4}} \put(65,0){$y_1$}
    \put(130,8){\line(0,1){4}} \put(125,0){$y_2$}
    \put(180,8){\line(0,1){4}} \put(175,0){$y_3$}
    \put(210,8){\line(0,1){4}} \put(203,0){$+1$}
    \put(10,10){\line(1,1){60}}
    \put(70,70){\circle*{4}} \put(65,78){$g_1$}
    \put(70,70){\line(6,1){60}}
    \put(130,80){\circle*{4}} \put(128,88){$g_2$}
    \put(130,80){\line(5,-2){50}}
    \put(180,60){\circle*{4}} \put(178,68){$g_3$}
    \put(180,60){\line(3,-5){30}}
    \end{picture}
  \end{center}
  \caption{Discrete Dirichlet string with three masses}\label{f1}
\end{figure}

The spectral problem for the discrete string can be conveniently
reformulated using the transition matrix associated with the initial
value problem posed at the left end $y=-1$.
One starts at $\phi(-1)=0$ with initial slope $\phi_y(-1)=1$,
and propagates the solution to the right end point,
which results in
\begin{equation}
  \label{eq:string-propagate}
  \begin{pmatrix}
    \phi(1;z)\\ \phi_y(1;z)
  \end{pmatrix}
  =L_n \, G_n(z) \, \cdots L_1 \, G_1(z) \, L_0
  \begin{pmatrix}
    0\\1
  \end{pmatrix},
\end{equation}
where
\begin{equation*}
  L_k=\begin{pmatrix}1&l_k\\0&1\end{pmatrix},
  \qquad
  G_k(z)=\begin{pmatrix}1&0\\-z g_k& 1\end{pmatrix}.
\end{equation*}
(Like for the cubic string, we set $l_k=y_{k+1}-y_k$, and also $k'=n+1-k$.)
The eigenvalues are the zeros of the $n$th degree polynomial
$\phi(1;z)$,
and consequently they are the poles of the \emph{Weyl function}
of the problem, defined by
\begin{equation}
  \label{eq:string-weyl}
  W(z)=\frac{\phi_y(1;z)}{\phi(1;z)}.
\end{equation}
With
\begin{equation}
  \label{eq:string-convergents}
  \begin{split}
    L_n &=
    \begin{pmatrix}
      Q_0 & Q_1 \\
      P_0 & P_1
    \end{pmatrix}
    =: a^{(1)},
    \\
    L_n \, G_n(z) &=
    \begin{pmatrix}
      Q_2 & Q_1 \\
      P_2 & P_1
    \end{pmatrix}
    =: a^{(2)},
    \\
    L_n \, G_n(z) \, L_{n-1} &=
    \begin{pmatrix}
      Q_2 & Q_3 \\
      P_2 & P_3
    \end{pmatrix}
    =: a^{(3)},
  \end{split}
\end{equation}
and so on,
equation \eqref{eq:string-propagate} is just a matrix version
of the standard recurrence for the \emph{convergents}
\begin{equation*}
  \frac{0}{1} = \frac{P_0}{Q_0},
  \quad
  \frac{1}{l_n} = \frac{P_1}{Q_1},
  \quad
  \cfrac{1}{l_n +
    \cfrac{1}{z g_n}} = \frac{P_2}{Q_2},
  \quad
  \cfrac{1}{l_n +
    \cfrac{1}{z g_n +
      \cfrac{1}{l_{n-1}}}}= \frac{P_3}{Q_3},
  \quad
  \dots
\end{equation*}
of a continued fraction expansion of Stieltjes type \cite{stieltjes},
\begin{equation}
  \label{eq:W-contfrac}
  W(z)
  = \cfrac{1}{l_n +
    \cfrac{1}{z g_n +
      \cfrac{1}{l_{n-1} +
        \cfrac{1}{\raisebox{1.5ex}{$\ddots$} +
          \cfrac{1}{z g_2 +
            \cfrac{1}{l_1+
              \cfrac{1}{z g_1 +
                \cfrac{1}{l_0}
                }}}}}}}
  = \frac{P_{n+1}}{Q_{n+1}}.
\end{equation}
The \emph{remainders}
\begin{equation*}
  R_0 = \frac{1}{l_0},
  \quad
  R_1 =
  \cfrac{1}{z g_1 +
    \cfrac{1}{l_0}},
  \quad
  R_2 =
  \cfrac{1}{l_1 +
    \cfrac{1}{z g_1 +
      \cfrac{1}{l_0}}},
  \quad
  \dots
\end{equation*}
in the continued fraction are given by
\begin{equation}
  \label{eq:string-remainders}
  R_{2k-1} = \frac{q_k}{p_k},
  \quad
  R_{2k} = \frac{p_k}{q_{k+1}},
  \quad\text{where}\quad
  \begin{pmatrix}
    q_k \\ p_k
  \end{pmatrix}
  =
  \begin{pmatrix}
    \phi(y_{k'};z) \\
    \phi_y(y_{k'}-;z)
  \end{pmatrix}.
\end{equation}
Note that the convergents $(P_j,Q_j)$ are obtained by formally setting
the remainder $R_j$ to zero in \eqref{eq:W-contfrac}.
Write
\begin{equation}
  \label{eq:W-parfrac}
  \frac{W(z)}{z} = \sum_{k=0}^n \frac{a_k}{z-\lambda_k}
  = \sum_{i=0}^{\infty} \frac{(-1)^j A_j}{z^{j+1}},
\end{equation}
where $\lambda_0=0$, $a_0=1/2$, and where
\begin{equation}
  A_j = \sum_{k=0}^n (-\lambda_k)^j a_j = \int x^j \, d\mu(x)
\end{equation}
is the $j$th moment of the \emph{spectral measure}
\begin{equation}
  \label{eq:string-mu}
  \mu = \sum_{k=0}^n a_k \delta_{-\lambda_k}.
\end{equation}
The convergents are Pad\'e approximants of $W$:
\begin{gather*}
  W(z) - \frac{P_{2k-1}(z)}{Q_{2k-1}(z)} = O\left(\frac{1}{z^{2k-1}}\right),
  \\
  \deg P_{2k-1}=k-1,
  \qquad
  \deg Q_{2k-1}=k-1,
  \qquad
  P_{2k-1}(0)=1,
\end{gather*}
in the odd case, and
\begin{gather*}
  W(z) - \frac{P_{2k}(z)}{Q_{2k}(z)} = O\left(\frac{1}{z^{2k}}\right),
  \\
  \deg P_{2k}=k-1,
  \qquad
  \deg Q_{2k}=k,
  \qquad
  Q_{2k}(0)=1,
\end{gather*}
in the even case.
These conditions uniquely determine the coefficients of $Q_j(z)$
in terms of certain Hankel determinants $\Delta_k^i$ involving
the moments $A_j$ appearing in the Laurent series of $W(z)$.
Since
\begin{equation*}
  \begin{split}
    Q_{2k-1}(z) &=
    \bigl( g_n \, l_n \, g_{n-1} \, l_{n-1} \cdots g_{k'} \bigr) z^{k-1} + \ldots,
    \\
    Q_{2k}(z) &=
    \bigl( g_n \, l_n \, g_{n-1} \, l_{n-1} \cdots g_{k'} \, l_{k'} \bigr) z^k + \ldots,
  \end{split}
\end{equation*}
we can express the string data as ratios of coefficients of $Q_j$'s,
which themselves are ratios of Hankel determinants,
resulting in altogether four determinants.
In the notation of \cite{bss-moment},
\begin{equation*}
  l_{k'} = \frac{\bigl( \Delta_{k-1}^1 \bigr)^2}{\Delta_{k-1}^0 \Delta_k^0},
  \qquad
  g_{k'} = \frac{\bigl( \Delta_{k}^0 \bigr)^2}{\Delta_{k-1}^1 \Delta_k^1}.
\end{equation*}
These Hankel determinants can be evaluated in terms of $\lambda_k$'s and
$a_k$'s using Heine's formula \eqref{eq:Hankel-Heine},
thereby recovering the string data $\{ g_i,l_i \}$
in terms of symmetric functions of the spectral data
$\{ \lambda_k, a_k \}$ encoded in the Weyl function.

\section{Appendix: Combinatorial identities}
\label{sec:app-identities}

In the proof of Lemma~\ref{lem:super-Heine} we used some combinatorial
identities for symmetric polynomials.

\begin{lemma}
  \label{lem:indentities}
  With the notation of Lemma~\ref{lem:super-Heine}
  as well as that of Definition~\ref{def:notation-galore} in force,
  the following identities hold.
  \begin{enumerate}
  \item 
    With the sums running over all partitions of $\{1,2,\dots,2k\}$
    into disjoint subsets $I$ and $J$ with $\abs{I}=\abs{J}=k$,
    \begin{equation}\label{eq:firstident}
      \sum_{I,J}\Delta_I^2 \Delta_J^2 \Gamma_I \Gamma_J x_I=
      \frac{1}{2^k} \sum_{I,J}\Delta_I^2 \Delta_J^2\Gamma_{I,J}.
    \end{equation}
  \item
    With the sums running over all partitions of $\{0,1,2,\dots,2k\}$
    into disjoint subsets $I$ and $J$ with
    $\abs{I}=k$ and $\abs{J}=k+1$,
    \begin{align}
      \sum_{I,J}\Delta_I^2 \Delta_J^2 \Gamma_I \Gamma_J x_I=&
      \frac{1}{2^k} \sum_{I,J}\Delta_I^2 \Delta_J^2\Gamma_{I,J},\\
      \sum_{I,J}\Delta_I^2 \Delta_J^2\Gamma_I \Gamma_J (x_I)^2=&
      \frac{1}{2^k} \sum_{I,J}\Delta_I^2 \Delta_J^2x_I \Gamma_{I,J}.
    \end{align}
  \end{enumerate}
\end{lemma}

\begin{proof}
  We give a detailed proof only for the first identity.  Almost
  identical technique can be used to verify the other identitities.
  We start by noting that both sides of \eqref{eq:firstident}
  are symmetric polynomials in the variables $x_1,x_2,\dots,x_{2k}$.
  In each single variable $x_a$ both sides are of degree $3k-2$.  Let
  us denote the difference of the two sides by $P$.
  We claim that
  \begin{enumerate}
  \item $P$ is divisible by $\Delta(x_1,x_2,x_3,\dots,x_{2k})$.
  \item $P$ is divisible by $\Delta^2(x_1,x_2,x_3,\dots,x_{2k})$.
  \end{enumerate}
  Granted Claim~2 we easily conclude that $P=0$ and
  hence the identity is proven.   Indeed,
  the degree of $P$ in, say, the $x_1$ variable is $3k-2$, while that
  of $\Delta^2(x_1,x_2,x_3,\dots,x_{2k})$ is $4k-2$.
  
  Furthermore, Claim~2 follows from Claim~1 due to the
  symmetry of $P$, the skew-symmetry of $\Delta(x_1,x_2,x_3,\dots,x_{2k})$,
  and the fact that any polynomial skew-symmetric with respect to
  all transpositions of variables
  (in particular, then, the polynomial $P/\Delta$)
  must be divisible by
  $\Delta(x_1,x_2,x_3,\dots, x_{2k})$.

  Thus it suffices to prove Claim~1.  Moreover,
  by symmetry, it suffices to show that $P$ is divisible by $x_1-x_2$.
  The proof proceeds by induction. The base case $k=0$ is just $1=1$.
  Assume the identity holds for $k=m-1$. We will be evaluating both sides
  of \eqref{eq:firstident} at $x_1=x_2$.  In view of this, the only
  nonvanishing contributions to
  the left-hand side are coming from sets $I$ and $J$ such that
  $I=\{1\}\cup I'$ and $J=\{2\}\cup J'$,
  or such that
  $I=\{2\}\cup I''$ and $J=\{1\}\cup J''$.
  Thus the only terms contributing to the left-hand side will be
  \begin{multline*}
      \sum_{I',J'}\Delta_{\{1\},I'}^2\Delta_{I'}^2 \Delta_{\{2\},J'}\Delta_{J'}^2
      \Gamma_{\{1\},I'}\Gamma_{I'} \Gamma_{\{2\},J'}\Gamma_{J'}x_1 x_{I'}
      \\
      +\sum_{I'',J''}\Delta_{\{2\},I''}^2\Delta_{I''}^2 \Delta_{\{1\},J''}
      \Delta_{J''}^2
      \Gamma_{\{2\},I''}\Gamma_{I''} \Gamma_{\{1\},J''}\Gamma_{J'}x_2 x_{I''}.
  \end{multline*}
  Since $I'\cup J'=I''\cup J''=\{3,4,\dots, 2k\}=:A$, we obtain after
  evaluating at $x_1=x_2$ that the above sums contribute
  \begin{equation*}
    2x_2\Delta_{\{2\},A}^2\Gamma_{\{2\},A}\sum_{I',J'}\Delta_{I'}^2 \Delta_{J'}^2
    \Gamma_{I'} \Gamma_{J'}x_{I'}.
  \end{equation*}
  Likewise, following the same steps, the right-hand side contributes
  \begin{equation*}
    \frac{4x_2}{2^k} \Delta_{\{2\},A}^2\Gamma_{\{2\},A}\sum_{I',J'}\Delta_{I'}^2
    \Delta_{J'}^2
    \Gamma_{I', J'}.
  \end{equation*}
  Since $\abs{I'}=\abs{J'}=k-1$, the induction hypothesis ensures that the
  two contributions are equal.
  Thus $P$ vanishes whenever $x_1=x_2$,
  so it is divisible by $x_1-x_2$. This proves Claim~1.
\end{proof}

\begin{remark}
  The only similar identity for symmetric polynomials
  that we have come across in the literature
  is one involving $\Delta(x)^2$, but not $\Gamma(x)$, in a paper
  by Anderson \cite{anderson} on the Toda lattice.
\end{remark}

\bibliographystyle{hplain}
\bibliography{DPcubic-arxiv}

\end{document}